\documentclass{lmcs}
\usepackage[utf8]{inputenc}
\pdfoutput=1

\usepackage{lastpage}
\lmcsdoi{19}{4}{20}
\lmcsheading{}{\pageref{LastPage}}{}{}%
{Oct.~26,~2021}{Dec.~07,~2023}{}


\usepackage{xspace,amstext,amssymb,url}

\usepackage{mathtools}

\usepackage[ruled,linesnumbered,vlined]{algorithm2e}

\SetKwComment{Comment}{$\vartriangleright$  }{}

\usepackage{tikz}
\usetikzlibrary{decorations.pathmorphing, automata, arrows, shapes,calc}%
\usetikzlibrary{decorations.pathreplacing}

\usepackage{hyperref} %

\newcommand{\cL}{{\ensuremath{L}}}
\newcommand{\nat}{\mathbb{N}}

\newcommand{\cC}{\mathcal{C}}

\newcommand{\lab}{\text{lab}}
\newcommand{\true}{\ensuremath{\mathsf{true}}\xspace}
\newcommand{\false}{\ensuremath{\mathsf{false}}\xspace}

\newcommand{\FO}{\ensuremath{\mathbf{FO}}\xspace}
\newcommand{\dc}{\ensuremath{\mathsf{DC}}\xspace}
\newcommand{\etract}{\ensuremath{\mathsf{T}_\mathsf{tract}}\xspace}
\newcommand{\ctract}{\ensuremath{\mathsf{SP}_\mathsf{tract}}\xspace}
\newcommand{\ttract}{\etract}
\newcommand{\sptract}{\ctract}
\newcommand{\rspq}{\textsf{RSPQ}\xspace}
\newcommand{\rtq}{\textsf{RTQ}\xspace}

\newcommand{\pedge}{\ensuremath{P_\mathsf{edge}}\xspace}
\newcommand{\Edge}{\ensuremath{\mathsf{Edge}}\xspace}
\newcommand{\Left}{\ensuremath{\mathsf{left}}\xspace}
\newcommand{\Right}{\ensuremath{\mathsf{right}}\xspace}
\newcommand{\alphabet}{\ensuremath{\operatorname{Alph}\xspace}}
\newcommand{\loopwords}{\ensuremath{\operatorname{Loop}\xspace}}
\newcommand{\Cuts}{\mathsf{Cuts}\xspace}
\newcommand{\component}{{component}\xspace}
\newcommand{\components}{{components}\xspace}

\newcommand{\edgedispaths}{\textsf{Two\-Edge\-Disjoint\-Paths}\xspace}

\newcommand{\lscp}{left-synchronized containment property\xspace}
\newcommand{\rscp}{right-synchronized containment property\xspace}
\newcommand{\lorrscp}{left- or right-synchronized containment property\xspace}
\newcommand{\lsla}{left-synchronized power abbreviations\xspace}
\newcommand{\rsla}{right-synchronized power abbreviations\xspace}
\newcommand{\lorrsla}{left- or right-synchronized power abbreviations\xspace}

\newcommand{\edge}{\mathcal{E}\xspace}
\newcommand{\orig}{\operatorname{origin}\xspace}
\newcommand{\dest}{\operatorname{destination}\xspace}

\newcommand{\np}{\text{NP}\xspace}
\newcommand{\pspace}{\text{PSPACE}\xspace}

\newcommand{\nlogspace}{\text{NL}\xspace}
\newcommand{\aczero}{\text{AC}\ensuremath{^0}\xspace}

\newcommand{\Abbrv}{\mathsf{Abbrv}}

\newcommand{\LoopSym}[1]{\Sigma^\circlearrowright(#1)}

\newlength\boxwidth%
\newlength\questionwidth%
\newcommand{\decisionproblem}[4]{
    \setlength\boxwidth{#1-.5cm}{
        \setlength\questionwidth{#1-.5cm}\addtolength\questionwidth{-2.5cm}{
            \begin{center}
                \fbox{\parbox[t]{\boxwidth}{\centerline{#2}
                        \vspace{0mm}
                        \begin{tabular}{l@{\hspace{5mm}}p{\questionwidth}}
                            Given: & #3\\[1pt]
                            Question: & #4\\
                \end{tabular}}}
            \end{center}
}}}

\begin{document}

		\title[A Trichotomy for Regular Trail Queries]{A Trichotomy for Regular Trail Queries}
\titlecomment{{\lsuper*}A shorter version of this article has appeared in STACS 2020, see~\cite{MartensNT-stacs20}.}

\author[W.~Martens]{Wim Martens\lmcsorcid{0000-0001-9480-3522}} 
\author[M.~Niewerth]{Matthias Niewerth\lmcsorcid{0000-0003-2032-5374}} 
\author[T.~Popp]{Tina Popp\lmcsorcid{0000-0001-6355-3815}} 

\address{University of Bayreuth, Germany}

\begin{abstract}
  Regular path queries (RPQs) are an essential component of graph query
  languages. Such queries consider a regular expression $r$ and a directed
  edge-labeled graph $G$ and search for paths in $G$ for which the sequence of
  labels is in the language of $r$. In order to avoid having to consider
  infinitely many paths, some database engines restrict such paths
  to be \emph{trails}, that is, they only consider paths without repeated edges.
  In this article we consider the evaluation problem for RPQs under trail
  semantics, in the case where the expression is fixed. We show that, in this
  setting, there exists a trichotomy. More precisely, the complexity of RPQ evaluation divides the regular languages into the
  finite languages, the class \ttract (for which the problem is
  tractable), and the rest. Interestingly, the tractable class in the trichotomy is larger
  than for the trichotomy for \emph{simple paths}, discovered by Bagan, Bonifati, and Groz [JCSS 2020]. In addition to this trichotomy result, we also
  study characterizations of the tractable class, its expressivity, the
  recognition problem, closure properties, and show how the decision problem can
  be extended to the enumeration problem, which is relevant to practice.
\end{abstract}

\maketitle

\section{Introduction}
Graph databases are a popular tool to model, store, and
analyze data~\cite{neo,tigergraph,oracle,wikidata,dbpedia}. They are engineered
to make the \emph{connectedness of data} easier to analyze. This is indeed a
desirable feature, since some of today's largest companies have become so
successful because they understood how to use the connectedness of the data in
their specific domain (e.g., Web search and social media). One aspect of graph
databases is to bring tools for analyzing connectedness to the masses.

Regular path queries (RPQs) are a crucial component of graph databases, because
they allow reasoning about arbitrarily long paths in the graph and, in
particular, paths that are longer than the size of the query. A regular path
query essentially consists of
a regular expression $r$ and is evaluated on a
graph database which, for the purpose of this article, we view as an edge-labeled
directed graph $G$. When evaluated, the RPQ $r$ searches for paths in $G$ for
which the sequence of labels is in the language of $r$. The return type of the
query varies: whereas most academic research on RPQs~\cite{MendelzonW-sicomp95,Barcelo-pods13,BarceloLR-pods11,LosemannM-tods13,ArenasCP-www12}
and SPARQL~\cite{sparql11} focus on the first and last node of matching paths,
Cypher~\cite{cypher9} returns the entire paths. G-Core, a recent proposal by
partners from industry and academia, sees paths as ``first-class citizens'' in
graph databases~\cite{gcore}.

In addition, there is a large variation on which types of paths are considered.
Popular options are \emph{all paths}, \emph{simple paths}, \emph{trails}, and
\emph{shortest paths}. Here, \emph{simple paths} are paths without repeated nodes and
\emph{trails} are paths without repeated edges. Academic research has focused mostly on \emph{all paths},
but Cypher 9~\cite{cypher9,FrancisGGLLMPRS-sigmod18}, which is perhaps the most
widespread graph database query language at the moment, uses \emph{trails}.
Since the trail semantics in graph databases has received virtually no attention
from the research community yet, it is crucial that we improve our understanding.

In this article, we study the \emph{data complexity} of RPQ evaluation under trail
semantics. That is, we study variants of RPQ evaluation in which the RPQ $r$ is
considered to be fixed. As such, the input of the problem only consists of an edge-labeled
(multi-)graph $G$ and a pair $(s,t)$ of nodes and we are asked if there exists a trail 
from $s$ to $t$ on which the sequence of labels matches $r$.
One of our main
results is a trichotomy on the RPQs for which this problem is in $\aczero$,
\nlogspace-complete, or \np-complete, respectively.
 By \ttract, we refer to the
class of tractable languages (assuming \np $\neq$ \nlogspace).

In order to increase our understanding of \ttract, we study several important
aspects of this class of languages. A first set of results is on
characterizations of \ttract in terms of closure properties and syntactic and
semantic conditions on their finite automata. In a second set of results, we therefore
compare the expressiveness of \ttract with yardstick languages such as
$\FO^2[<]$, $\FO^2[<,+1]$, $\FO[<]$ (or \emph{aperiodic languages}), and
\sptract. The latter class, \sptract, is the closely related class of languages
for which the data complexity of RPQ evaluation under \emph{simple path}
semantics is tractable.\footnote{Bagan et al.~\cite{BaganBG-jcss20} called the
  class \textsf{C}$_\text{tract}$, which stands for ``tractable class''. We
  distinguish between \ctract and \ttract here to avoid confusion between simple
  paths and trails.} Interestingly, \ttract is strictly larger than \sptract and
includes languages outside \sptract such as $a^*bc^*$ and $(ab)^*$ that are
relevant in application scenarios in network problems, genomic datasets, and
tracking provenance information of food products~\cite{PetraSelmerPersonal} and
were recently discovered to appear in public query
logs~\cite{BonifatiMT-vldb17,BonifatiMT-www19}. %
Furthermore, every
\emph{single-occurrence regular expression}~\cite{BexNSV-tods10} is in \ttract, which can be
a convenient guideline for users of graph databases, since
\emph{single-occurrence} (every alphabet symbol occurs at most once) is a very
simple syntactical property. It is also popular in practice:  we
analyzed the 50 million RPQs found in the logs
of~\cite{BonifatiMT-www18} and discovered that over 99.8\% of the RPQs are
single-occurrence regular expressions.

We then study the recognition problem for \ttract, that is: given an automaton,
does its language belong to \ttract? This problem is \nlogspace-complete (resp.,
\pspace-complete) if the input automaton is a DFA (resp., NFA). We also treat
closure under common operations such as union, intersection, reversal, quotients
and morphisms.

We conclude by showing that also the \emph{enumeration problem} is tractable
for \ttract. By tractable, we mean that the paths that match the RPQ can be
enumerated with only \emph{polynomial delay} between answers. Technically, this
means that we have to prove that we cannot only solve a decision variant of the
RPQ evaluation problem, but we also need to find witnessing paths. We prove that the
algorithms for the decision problems can be extended to return \emph{shortest
  paths}. This insight can be combined with Yen's Algorithm~\cite{Yen-ms71} to give a polynomial
delay enumeration algorithm.

\smallskip
\noindent \textbf{Related Work.} RPQs on graph databases have been studied since the
end of the 80's and are now finding their way into commercial products. The literature usually considers the variant of RPQ
evaluation where one is given a graph database $G$, nodes $s,t$, and an RPQ $r$,
and then needs to decide if $G$ has a path from $s$ to $t$ (possibly with loops) that matches
$r$. For arbitrary and shortest paths, this problem is well-known to be
tractable, since it boils down to testing intersection emptiness of two NFAs.

Mendelzon and Wood~\cite{MendelzonW-sicomp95} studied the problem for simple
paths, which are paths without node repetitions. They observed that the problem
is already NP-complete for regular expressions $a^*ba^*$ and $(aa)^*$. These two
results rely heavily on the work of Fortune et al.~\cite{FortuneHW-TCS80} and
LaPaugh and Papadimitriou~\cite{LapaughP-networks84}.

Our work is most closely related to the work of Bagan et
al.~\cite{BaganBG-jcss20} who, like us, studied the complexity of RPQ evaluation
where the RPQ is fixed. They proved a trichotomy for the case where the RPQ
should only match simple paths. In this article we will refer to this class as
\ctract, since it contains the languages for which the \emph{simple path}
problem is tractable, whereas we are interested in a class for \emph{trails}.
Martens and Trautner~\cite{MartensT-tods19} refined this trichotomy of Bagan et
al.~\cite{BaganBG-jcss20} for \emph{simple transitive expressions}, by
analyzing the complexity where the input consists of both the expression and the graph.

Paperman has integrated the classes \sptract and \ttract in his tool called
Semigroup Online~\cite{PapermanTool}. The tool can process a regular expression
as input and can tell the user whether the language is in \sptract, \ttract, and/or
in many other important classes of languages.

\smallskip
\noindent \textbf{Trails versus Simple Paths.}
We conclude with a note on the relationship between simple paths and trails.
For many computational problems, the complexities of dealing with simple paths or trails are
the same due to two simple reductions, namely: (1) constructing the line graph or
(2) splitting each node into two, see for example Perl and Shiloach~\cite[Theorem 2.1
and 2.2]{PerlS-jacm78}.
As soon as we consider labeled graphs, the line graph technique still works, but
not the nodes-splitting technique, because the labels on paths change. As a
consequence, we know that finding trails is at most as hard as finding simple
paths, but we do not know if it has the same complexity when we require that they match a certain RPQ $r$.

In this article we show that the relationship is strict, assuming \nlogspace
$\neq$ \np. An easy example is the
language $(ab)^*$, which is NP-hard for simple paths~\cite{LapaughP-networks84,
  MendelzonW-sicomp95}, but---assuming that $a$-labeled edges are different from $b$-labeled edges---in
\nlogspace for trails. This is because every path from $s$ to $t$ that matches $(ab)^*$
can be reduced to a trail from $s$ to $t$ that matches $(ab)^*$ by removing
loops (in the path, not in the graph) that match $(ab)^*$ or $(ba)^*$. In Figure~\ref{fig:ab*} we depict four
small graphs, all of which have trails from $s$ to $t$. (In the three rightmost graphs,
there is exactly one path labeled $(ab)^*$, which is also a trail.)

\begin{figure}[t] %
    \centering {
    \begin{tikzpicture}[auto,>=latex,->]%
\node (s) at (0,0) {};
\node (v1) at (1,0) {};
\node (v2) at (1,1) {};
\node (t) at (2,0) {};

\foreach \i in {s,v1,v2,t}{
    \fill (\i) circle (2pt);
}

\draw (s) coordinate[label={left:$s$}];
\draw (t) coordinate[label={right:$t$}];

\path (s) edge node {$a$} (v1);
\path (v1) edge [bend left = 20] node  [yshift =5.8pt] {$b$} (v2);
\path (v2) edge [bend left = 20] node [yshift =5pt]  {$a$} (v1);
\path (v1) edge  node {$b$} (t);
        \end{tikzpicture}
    }
    \hspace{0.1cm}
     {\begin{tikzpicture}[auto,>=latex,->]
        \node (s) at (0,0) {};
        \node (v1) at (1,0) {};
        \node (t) at (2,0) {};

        \foreach \i in {s,v1,t}{
            \fill (\i) circle (2pt);
        }

        \draw (s) coordinate[label={left:$s$}];
        \draw (t) coordinate[label={right:$t$}];

        \path (s) edge node {$a$} (v1);
        \path (v1) edge [bend left = 20] node {$b$} (t);
        \path (t) edge [bend left = 20] node {$a$} (v1);
        \end{tikzpicture}
    }
     \hspace{0.1cm}
    {\begin{tikzpicture}[auto,>=latex,->]
        \node (s) at (0,0) {};
        \node (t) at (1.2,0) {};
        \node (v1) at (.6,.9) {};

        \foreach \i in {s,v1,t}{
            \fill (\i) circle (2pt);
        }

        \draw (s) coordinate[label={left:$s$}];
        \draw (t) coordinate[label={right:$t$}];

        \path (s) edge [bend right = 20, swap] node {$b$} (t);
        \path (s) edge [bend left = 20] node {$a$} (t);
        \path (t) edge [bend right = 20,swap] node {$b$} (v1);
        \path (v1) edge [bend right = 20,swap] node {$a$} (s);
        \end{tikzpicture}
    }
    \hspace{0.1cm}
     {\begin{tikzpicture}[auto,>=latex,->]
        \node (s) at (-.4,.9) {};
        \node (v1) at (0,0) {};
        \node (v2) at (1.2,0) {};
        \node (v3) at (.6,.9) {};
        \node (t) at (1.6,.9) {};
        \node (t2) at (2.4,.9) {};

        \foreach \i in {s,v1,v2,v3,t,t2}{
           \fill (\i) circle (2pt);
        }

        \draw (s) coordinate[label={left:$s$}];
        \draw (t2) coordinate[label={right:$t$}];

         \path (s) edge node {$a$} (v3);
         \path (v3) edge node {$a$} (t);
        \path (t) edge node {$b$} (t2);

        \path (v1) edge [bend left = 20] node  [yshift =-5pt]{$b$} (v3);
        \path (v3) edge [bend left = 20] node [yshift =-5pt]{$b$} (v2);
        \path (v2) edge [bend left = 20] node {$a$} (v1);
        \end{tikzpicture}
    }
    \caption{Directed, edge-labeled graphs that have a trail from $s$ to $t$.
    }%
\label{fig:ab*}
\end{figure}

\smallskip
\noindent \textbf{Outline.} We note that this is a full version of the work
presented in~\cite{MartensNT-stacs20}. In addition to adding the full proofs, we
generalize our results to multigraphs throughout the article.
In Section~\ref{sec:preliminaries} we define our notation, Section~\ref{sec:tractableclass} introduces the class $\ttract$, which contains exactly the regular languages for which finding a trail from $s$ to $t$ is in polynomial time (assuming $P \neq NP$). We prove this dichotomy in Section~\ref{sec:trichotomy}. (The article is named trichotomy because we can also differentiate between finite and infinite languages. For the first, finding such a path is in \aczero while NL-hard for the latter.) After giving some interesting closure properties of $\ttract$ in Section~\ref{sec:closureproperties} and extending the algorithm for languages in $\ttract$ to an enumeration algorithm, we conclude our work in Section~\ref{sec:conclusion}. The most complex part of this article is in Section~\ref{sec:tractableclass}, where we give several equivalent definitions of $\ttract$, some of which are needed for the proof of its tractability, others, like the syntactic definition given in Theorem~\ref{theo:classification} might be useful for database engineers, while others are used to compare $\ttract$ to well-known classes such as \textbf{FO} or $\text{\textbf{FO}}^2[<,+1]$.

\section{Preliminaries}\label{sec:preliminaries}
We use $[n]$ to denote the set of integers $\{1,\ldots, n\}$.
By $\Sigma$ we
always denote a finite alphabet, i.e., a finite set of \emph{symbols}. We always
 denote symbols by $a$, $b$, $c$, $d$ and their variants, like $a'$, $a_1$, $b_1$, etc.
 The regular expressions we use in this article are defined as follows: $\emptyset, \varepsilon$, and every symbol in $\Sigma$ is a regular expression. When $r$ and $s$ are regular expressions, then $(rs)$, $(r+s)$, $(r?)$, $(r^*)$, and $(r^+)$ are also regular expressions. We use the usual precedence rules to omit parentheses. For $n \in \nat$, we use $r^n$ to abbreviate the $n$-fold concatenation $r\cdots r$ of $r$. The language $L(r)$ of a regular expression $r$ is defined as usual. For readability, we often omit the $L(\cdot)$ and only write $r$ for the language of $r$.
A \emph{word} is a finite sequence $w = a_1 \cdots a_n$ of symbols.

We consider edge-labeled directed multigraphs $G = (V,E,\edge)$, where $V$ is a
finite set of nodes, $E$ is a finite set of edges, and
$\edge \colon E \to V \times \Sigma \times V$ is a function that maps each edge
identifier to a tuple ($v_1,a,v_2)$ describing the origin, the label, and the
destination node of the edge. We denote $v_1$ by $\orig(e)$, $a$ by $\lab(e)$
and $v_2$ by $\dest(e)$. We emphasize that $\edge$ does not need to be
injective, i.e., there might be several edges with identical origin, label, and
destination. The size of $G$ is defined as
$|V|+|E|$. A (simple) graph is a multigraph where $\edge$ is injective. A
\emph{path} $p$ from node $s$ to $t$ is a sequence $e_1 \cdots e_m$ of edges
such that $\orig(e_1)=s$, $\dest(e_m)=t$, and for $1 \leq i < m$ it holds that
$\dest(i)=\orig(i+1)$. By $|p|$ we denote the number of edges of a path. A path
is a \emph{trail} if every edge $e$ appears at most once\footnote{We note that
  it is allowed that for $i \neq j$ it holds that $\edge(e_i)=\edge(e_j)$.} and
a \emph{simple path} if all the nodes in $\orig(e_1)$ and
$\dest(e_1),\dots,\dest(e_m)$ are different. We note that each simple path is a
trail but not vice versa. We denote $\lab(e_1) \cdots \lab(e_m)$ by $\lab{(p)}$.
Given a language $L \subseteq \Sigma^*$, path $p$ \emph{matches $L$} if
$\lab(p) \in L$. For a subset $E' \subseteq E$, path $p$ is $E'$-restricted if
every edge of $p$ is in $E'$. Given a trail $p$ and two edges $e_1$ and $e_2$ in
$p$, we denote the subpath of $p$ from $e_1$ to $e_2$ by $p[e_1,e_2]$.

We define an NFA $A$ to be a tuple $(Q,\Sigma,I,F,\delta)$
where $Q$ is the finite set of states; $I \subseteq  Q$
is a set of initial states; $\delta
\subseteq Q \times \Sigma \times Q$ is
the transition relation; and $F \subseteq Q$ is the set of accepting states.
 Strongly connected components of (the graph of) $A$ are simply called
\emph{components}. Unless noted otherwise, \components will be non-trivial, i.e., containing at least one edge.
We write $C(q)$ to denote the strongly connected component of state $q$.

By $\delta(q,w)$ we denote the states reachable from state $q$ by reading $w$. Given a path $p$, we also slightly abuse notation and write $\delta(q,p)$ instead of $\delta(q,\lab(p))$.
We denote by $q_1 \leadsto q_2$ that state $q_2$ is reachable from $q_1$.
Finally, $\cL_{q}$ denotes the set of all words accepted from $q$ and $L(A) = \bigcup_{q \in I} \cL_q$ is the set of words accepted by $A$.
 For every state $q$, we denote by $\loopwords(q)$ the set $\{w \in \Sigma^+ \mid
\delta_L(q,w)=q\}$ of all non-empty words
that allow to loop on $q$. For a word $w$ and a language $L$, we define $wL =
\{w w' \mid w' \in L\}$ and $w^{-1}L=\{w' \mid ww' \in L\}$.

A DFA is an NFA such that $I$ is a singleton and for all $q \in Q$ and $\sigma \in \Sigma$: $|\delta(q,\sigma)| \leq 1$.
Let $L$ be a regular language. We denote by $A_L
= (Q_L,\Sigma, i_L, F_L, \delta_L)$
the (complete) minimal DFA for $L$ and by
$N$ the number $|Q_L|$ of states.
For $q_0 \in Q$, we say that a \emph{run from $q_0$} of $A$ on $w = a_1 \cdots a_n$ is a sequence
$q_0 \rightarrow \cdots \rightarrow q_n$ of states such that $q_i \in \delta(q_{i-1},a_i)$, for every $i
\in \{1,\ldots,n\}$. When $A$ is a DFA and $q_0$ its initial state, we also simply call
it  \emph{the run of $A$ on $w$}.
The product of multigraph $G=(V,E,\edge)$ and NFA $A=(Q,\Sigma,I,F,\delta)$ is a graph $(V',E',\edge')$ with $V' = V \times Q$, $E' = \{(e,(q_1,q_2)) \mid (q_1,\lab(e),q_2)\in \delta\}$ and $\edge'((e,(q_1,q_2)))=((\orig(e),q_1),\lab(e),(\dest(e),q_2))$.

A language $L$ is \emph{aperiodic} if and only if $\delta_L(q, w^{N+1}) =
\delta_L(q,w^{N})$ for every state $q$ and word $w$.
Equivalently, $L$ is aperiodic if and only if its minimal DFA does not have simple cycles labeled $w^k$ for $k>1$ and $w \neq \varepsilon$. Thus, for ``large enough $n$'' we have: $u w^n v \in L$ iff $u w^{n+1} v \in L$. So, a language like $(aa)^*$ is not aperiodic (take $w=a$ and $k=2$), but $(ab)^*$ is.
(There are many
characterizations of aperiodic languages~\cite{Schutzenberger-iandc65}.)

We study the
\emph{regular trail query (\rtq) problem} for a regular language $L$.

\decisionproblem{.75\linewidth}{$\rtq(L)$}{A (multi-)graph $G = (V,E, \edge)$ and $(s,t) \in 
  V\times V$.}{Is there a trail from $s$ to $t$ that matches $L$?}
A similar problem, which was studied by  Bagan et al.~\cite{BaganBG-jcss20}, is
the \emph{$\rspq$ problem}. The $\rspq(L)$ problem asks if there exists a \emph{simple path}
from $s$ to $t$ that matches $L$. %

\section{The Tractable Class}\label{sec:tractableclass}

In this section, we define and characterize a class of languages of which we
will prove that it is exactly the class of regular languages $L$ for which
$\rtq(L)$ is tractable (if NL $\neq$ NP).

\subsection{Warm-Up: Downward Closed Languages}

It is instructive to first discuss the case of downward closed languages. A
language $L$ is \emph{downward closed} (\dc) if it is closed under taking
subsequences. That is, for every word $w=a_1 \cdots a_n \in L$ and every
sequence $0<i_1 < \cdots <i_k<n+1$ of integers, we have that $a_{i_1} \cdots
a_{i_k} \in L$. Perhaps surprisingly, \emph{downward closed languages are always
  regular}~\cite{Haines-jct69}. Furthermore, they can be defined by a clean class of regular
expressions (which was shown by Jullien~\cite{Jullien-phd69} and later rediscovered by
Abdulla et al.~\cite{AbdullaCBJ-fmsd04}), which is defined as follows.

\begin{defi}
  An \emph{atomic expression} over $\Sigma$ is an expression of the form
  $(a+\varepsilon)$ or of the form $(a_1 + \cdots + a_n)^*$, where
  $a,a_1,\ldots,a_n \in \Sigma$. A \emph{product} is a (possibly empty)
  concatenation $e_1 \cdots e_n$ of atomic expressions $e_1,\ldots,e_n$. A
  \emph{simple regular expression} is of the form $p_1 + \cdots + p_n$, where
  $p_1,\ldots,p_n$ are products.
\end{defi}
Another characterization is by Mendelzon and Wood~\cite{MendelzonW-sicomp95}, who show that a
regular language $L$ is downward closed if and only if its minimal DFA $A_L=
(Q_L,\Sigma,i_L,F_L,\delta_L)$ exhibits the \emph{suffix language containment property},
which says that if $\delta_L(q_1, a) = q_2$ for some symbol $a \in \Sigma$,
then we have $\cL_{q_2} \subseteq \cL_{q_1}$.\footnote{They restrict $q_1, q_2$ to be on paths from $i_L$ to some state in $F_L$, but the property trivially holds for $q_2$ being a sink-state.}
Since this property is transitive, it is equivalent to require that $\cL_{q_2} \subseteq
\cL_{q_1}$ for every state $q_2$ that is reachable from $q_1$.
\begin{thmC}[\cite{AbdullaCBJ-fmsd04,Haines-jct69,Jullien-phd69,MendelzonW-sicomp95}]\label{theo:downwardclosed}
  The following are equivalent:
  \begin{enumerate}[(1)]
  \item $L$ is a downward closed language.
  \item $L$ is definable by a simple regular expression.
  \item The minimal DFA of $L$ exhibits the suffix language containment property.
  \end{enumerate}
\end{thmC}

\noindent
Obviously, $\rtq(L)$ is tractable for every downward closed language $L$, since
it is equivalent to deciding if there exists a path from $s$ to $t$ that matches
$L$. For the same reason, deciding if there is a \emph{simple path} from $s$ to
$t$ that matches $L$ is also tractable for downward closed languages.
However, there are languages that are not downward closed for which we show
$\rtq(L)$ to be
tractable, such as $a^*bc^*$ and $(ab)^*$. %
For these two languages, the simple path
variant of the problem is intractable.

\subsection{Main Definitions and Equivalence}

The following definitions are the basis of the class of languages
for which
$\rtq(L)$ is tractable.

\begin{defi}\label{def:leftsync}
  An NFA $A$ satisfies the \emph{left-synchronized containment property}
  if there exists an $n \in \nat$ such that the following implication holds for all $q_1, q_2 \in Q$ and $a \in \Sigma$:
 \begin{multline*}
	 \text{ If } q_1 \leadsto q_2 \text{ and if } w_1 \in
	  \loopwords(q_1),  w_2 \in \loopwords(q_2)
	  \text{ with } w_1 = a w_1' \text{  and } w_2 = aw_2'\text{,} \\
	  \text{ then  } w_2^n \cL_{q_2} \subseteq \cL_{q_1}.
  \end{multline*}
  Similarly, $A$ satisfies the \emph{right-synchronized containment
    property} if the same condition holds with $w_1 = w_1' a$ and
  $w_2 = w_2' a$.
\end{defi}
We illustrate this definition in Figure~\ref{fig:example-lscp}.
We note that the minimal DFA of any downward closed language satisfies the \lscp.

\begin{figure}[tbh] \centering
\begin{minipage}{.45\textwidth}
	\begin{tikzpicture}[->,>=latex,shorten >=.5pt,auto,node distance=1.3cm,
		inner sep = 0.7mm, initial text = {}]

		\node[state,initial,minimum size=17pt] (q1) at (-3,0) {$q_1$};
		\node[state,minimum size=17pt] (q2) at (-3.7,1) {$q_2$};
		\node[state,minimum size=17pt] (q3) at (-2.3,1) {$q_3$};
		\node[state,minimum size=17pt] (q4) at (-1.5,0) {$q_4$};
		\node[state,minimum size=17pt] (q5) at (0,0) {$q_5$};
		\node[accepting,state,minimum size=17pt] (q6) at (1.5,0) {$q_6$};

		\path (q1) edge [bend left = 20] node {$a$} (q2);
		\path (q2) edge [bend left = 20] node {$b$} (q3);
		\path (q3) edge [bend left = 20] node {$c$} (q1);

		\path (q1) edge [swap] node {$a$} (q4);
		\path (q4) edge [swap] node {$b$} (q5);

		\path (q5) edge [bend left = 20] node {$c$} (q6);
		\path (q6) edge [bend left = 20] node {$a$} (q5);

	\end{tikzpicture}
\end{minipage}
\hspace{.2cm}
\begin{minipage}{.52\textwidth}\centering
	\begin{tikzpicture}[->,>=latex,shorten >=.5pt,auto,node distance=1.3cm,
	inner sep = 0.7mm, initial text = {}]

		\node[state,initial,minimum size=17pt] (q1) at (-3,0) {$q_7$};
		\node[state,minimum size=17pt] (q2) at (-3.7,1) {$q_8$};
		\node[state,minimum size=17pt] (q3) at (-2.3,1) {$q_{9}$};
		\node[state,minimum size=17pt] (q4) at (-1.5,0) {$q_{10}$};
		\node[state,minimum size=17pt] (q5) at (0,0) {$q_{11}$};
		\node[state,minimum size=17pt] (q6) at (1.5,0) {$q_{12}$};
		\node[accepting,state,minimum size=17pt] (q7) at (3,0) {$q_{13}$};

		\path (q1) edge [bend left = 20] node {$a$} (q2);
		\path (q2) edge [bend left = 20] node {$b$} (q3);
		\path (q3) edge [bend left = 20] node {$c$} (q1);

		\path (q1) edge [swap] node {$a$} (q4);
		\path (q4) edge [swap] node {$c$} (q5);
		\path (q5) edge [swap] node {$a$} (q6);

		\path (q6) edge [bend left = 20] node {$c$} (q7);
		\path (q7) edge [bend left = 20] node {$a$} (q6);

\end{tikzpicture}
\end{minipage}
	\caption{Example illustrating Definition~\ref{def:leftsync}. The left NFA does not satisfy the \lscp as $(ac)^* L_{q_6} \cap L_{q_1} = \emptyset$. The right NFA satisfies the \lscp with $n=2$ as $(ac)^2 L_{q_{13}} \subseteq L_{q_7}$ and $(ca)^2 L_{q_{12}}\subseteq L_{q_{9}}$.  }%
	\label{fig:example-lscp}
\end{figure}

  The \emph{left-synchronizing
  length} of an NFA $A$ is the smallest value $n$ such that the implication in
Definition~\ref{def:leftsync} for the left-synchronized containment property
holds. We define the \emph{right-synchronizing length} analogously.

\begin{obs}\label{obs:leftsync}
  Let $n_0$ be the left-synchronizing length of an NFA $A$. Then the implication
  of Definition~\ref{def:leftsync} is satisfied for every $n \geq n_0$. The
  reason is that $w_2 \in \loopwords(q_2)$.
\end{obs}

\begin{defi}\label{def:leftsyncpower}
  A regular language $L$ is \emph{closed under left-synchronized power
    abbreviations} (resp., \emph{closed under right-synchronized power
    abbreviations}) if there exists an $n \in \nat$ such that for all words
  $w_\ell, w_m, w_r \in \Sigma^*$ and all words $w_1 = a w_1'$ and $w_2 = a
  w_2'$ (resp., $w_1 = w_1' a$ and $w_2 = w_2' a$) we have that $w_\ell w_1^n
  w_m w_2^n w_r \in L$ implies $w_\ell w_1^n w_2^n w_r \in L$.
\end{defi}

We note that Definition~\ref{def:leftsyncpower} is equivalent to requiring that there exists an $n \in \nat$ such
that the implication holds for all $i \geq n$.
 The reason is that, given $i > n$ and a word of the form $w_\ell w_1^i w_m w_2^i
w_r$, we can write it as $w'_\ell w_1^n w_m w_2^n w'_r$ with $w'_\ell = w_\ell
w_1^{i-n}$ and $w'_r = w_2^{i-n} w_r$, for which the implication holds by
Definition~\ref{def:leftsyncpower}. %

  \begin{lem}\label{lem:n<N}
    Consider a minimal DFA $A_L = (Q_L,\Sigma,i_L, F_L,\delta_L)$ with $N$
    states. Then the following is true:
    \begin{enumerate}[(1)]
    \item If $A_L$ satisfies the \lscp, then the left-synchronizing length is at most $N$.
    \item If $A_L$ satisfies the \rscp, then the right-synchronizing length is at most $N$.
    \end{enumerate}
  \end{lem}
  \begin{proof}
    We only prove (1), (2) is symmetric. By Definition~\ref{def:leftsync}, there exists an $n\in \nat$ such
    that: If $q_1,q_2 \in Q_A$ and $a\in\Sigma$ such that $q_1 \leadsto q_2$ and if $w_1 \in
    \loopwords(q_1), w_2 \in \loopwords(q_2)$ with $w_1 = a w_1'$ and $w_2 = a
    w_2'$, then $w_2^n \cL_{q_2} \subseteq \cL_{q_1}$.

    If $n>N$, then there must be a loop in the $w_2^n$ part that generates multiples of $w_2$.
    Applying the pigeonhole principle there is an $i<n$ for which $w_2^i \cL_{q_2} \subseteq \cL_{q_1}$ holds.
    By repetition, we obtain an $i$ with $i <N$.
  \end{proof}

From Definition~\ref{def:leftsync}, Observation~\ref{obs:leftsync}, and Lemma~\ref{lem:n<N}, we get the following corollary.
\begin{cor}\label{cor:leftsync}
  Let $A$ be a minimal DFA with $N$ states, $q_1,q_2 \in Q_A$ with $q_1 \leadsto q_2$, $w_1 \in \loopwords(q_1)$, and $w_2 \in \loopwords(q_2)$. If  $A$ satisfies the
  \begin{itemize}
    \item \lscp, $w_1 = a w_1'$, and $w_2 = a w_2'$, then $w_2^N \cL_{q_2} \subseteq \cL_{q_1}$.
    \item \rscp, $w_1 = w_1' a$, and $w_2 = w_2'a$, then $w_2^N \cL_{q_2} \subseteq \cL_{q_1}$.
    \end{itemize}
\end{cor}

\noindent
We need two lemmas to prove Theorem~\ref{theo:equivalence}. And their proofs require the following lemma:
\begin{lem}[Implicit in~\cite{BaganBG-jcss20}, Lemma~3 proof]\label{lem:p}
	Every minimal DFA satisfying

	\vspace{-6mm}
	\begin{equation}\label{eq:loop}
		\text{ for all } q_1, q_2 \in Q_L \text{ such that }
		q_1 \leadsto q_2 \text{ and } \loopwords(q_1)\cap
		\loopwords(q_2)\neq \emptyset:  \cL_{q_2}\subseteq \cL_{q_1} \tag{P}
	\end{equation}
	\vspace{-6mm}

	\noindent accepts an aperiodic language.
\end{lem}

\begin{lem}\label{lem:rscp-aperiodic}\label{lem:lscp-aperiodic}
  If $A_L$ has the \lscp or \rscp, then $L$ is aperiodic.
\end{lem}
\begin{proof}
  Let $A_L$ satisfy the \lorrscp.
  We show that $L$ satisfies Property~\eqref{eq:loop}, restated here for convenience.
  \begin{equation}
    \cL_{q_2}\subseteq \cL_{q_1} \text{ for all } q_1, q_2 \in Q_L \text{ such that }
    q_1 \leadsto q_2 \text{ and } \loopwords(q_1)\cap \loopwords(q_2)\neq \emptyset \tag{\ref{eq:loop}}
  \end{equation}
  This proves the lemma since all languages satisfying Property~\eqref{eq:loop}
  are aperiodic, see Lemma~\ref{lem:p}.
  Let $q_1, q_2 \in Q_L$ and $w$ satisfy $q_1 \leadsto q_2$ and $w \in
  \loopwords(q_1) \allowbreak \cap \loopwords(q_2)$. By Corollary~\ref{cor:leftsync} we
  then have that $w^N \cL_{q_2}\subseteq \cL_{q_1}$. Since $w \in \loopwords(q_1)$, we have
  that $\delta(q_1,w^N) = q_1$, which in turn implies that $\cL_{q_2}\subseteq \cL_{q_1}$.
\end{proof}

\begin{lem}\label{lem:lsla-aperiodic}\label{lem:rsla-aperiodic}
  If $L$ is closed under \lorrsla, then $L$ is
  aperiodic.
\end{lem}
\begin{proof}
    Let $L$ be closed under \lorrsla and $i\in \nat$ be as in
    Definition~\ref{def:leftsyncpower}.
    We show that $A_L$ satisfies the Property~\eqref{eq:loop}.
    The aperiodicity then follows from Lemma~\ref{lem:p}.

     Let $q_1, q_2 \in Q_L$ and $w$ satisfy $q_1 \leadsto q_2$ and $w \in
    \loopwords(q_1) \allowbreak \cap \loopwords(q_2)$.
    Let $w_\ell, w_m \in \Sigma^*$ be such that $q_1 = \delta_L(i_L, w_\ell)$ and
    $q_2 = \delta_L(q_1,w_m)$. Let $w_r \in \cL_{q_2}$.
    Then, $w_\ell w^* w_m w^* w_r \subseteq L$ by construction.
    Especially, $w_\ell w^i w_m w^i w_r \in L$ and, by Definition~\ref{def:leftsyncpower},  also
    $w_\ell w^i w^i w_r \in L$. Since
    $\delta_L(i_L, w_\ell w^i w^i) = q_1$, this means that $w_r \in \cL_{q_1}$.
    Therefore, $\cL_{q_2} \subseteq \cL_{q_1}$.
\end{proof}

Next, we show that all conditions defined in Definitions~\ref{def:leftsync}
and~\ref{def:leftsyncpower} are equivalent for DFAs.
\begin{thm}\label{theo:equivalence}
  For a regular language $L$ with minimal DFA $A_L$, the following are equivalent:
  \begin{enumerate}[(1)]
  \item $A_L$ satisfies the left-synchronized containment property.
  \item $A_L$ satisfies the right-synchronized containment property.
  \item $L$ is closed under left-synchronized power abbreviations.
  \item $L$ is closed under right-synchronized power abbreviations.
  \end{enumerate}
\end{thm}
\begin{proof}
  Let $A_L = (Q_L,\Sigma,i_L,F_L,\delta_L)$.
  (1) $\Rightarrow$ (3): Let $A_L$ satisfy the \lscp. We will show that if there
  exists a word $w_\ell w_1^i w_m w_2^i w_r \in L$ with $i = N+N^2$ and $w_1$
  and $w_2$ starting with the same letter, then $w_\ell w_1^i w_2^i w_r \in L$.
  To this end, let $w_\ell w_1^i w_m w_2^i w_r \in L$. Due to the pumping lemma, there are
  states $q_1, q_2$ and integers $h,j,k,\ell,m,n \leq N$ with $j,m \geq 1$
  satisfying: $q_1 = \delta(i_L,w_\ell w_1^h)$, $q_1 = \delta(q_1, w_1^j)$,
  $q_2 = \delta(q_1, w_1^k w_m w_2^\ell)$, $q_2 = \delta(q_2, w_2^m)$, and
  $w_2^n w_r \in \cL_{q_2}$.  This implies that
  \[w_\ell w_1^h (w_1^j)^* w_1^k w_m w_2^\ell (w_2^m )^* w_2^n w_r \subseteq
    L\;.\] Since $A_L$ satisfies the \lscp and by Corollary~\ref{cor:leftsync},
  we have $(w_2^m)^N \cL_{q_2} \subseteq \cL_{q_1}$ and therefore
  \[w_\ell w_1^h (w_1^j)^* (w_2^m )^N w_2^n w_r \subseteq L\;.\] Now we use that
  $L$ is aperiodic, see Lemma~\ref{lem:lscp-aperiodic}:
  \[w_\ell w_1^h (w_1^j)^N (w_1)^* (w_2^m )^N(w_2)^* w_2^n w_r \subseteq L\] And
  finally, we use that $i = N+N^2$ and $h,j,m,n \leq N$
  to
  obtain $w_\ell (w_1)^i (w_2)^i w_r \in L$.

  (3) $\Rightarrow$ (4): Let $L$ be closed under \lsla and let $j \in \nat$ be
  the maximum of $|A_L|$ and $n+1$, where the $n$ is from
  Definition~\ref{def:leftsyncpower}. We will show that if $w_\ell (w_1 a)^j w_m
  (w_2 a)^j w_r \in L$, then $w_\ell (w_1 a)^j (w_2 a)^j w_r \in L$. If $w_\ell
  (w_1 a)^j w_m (w_2 a)^j w_r \in L$, then we also have $w_\ell (w_1 a)^j w_m
  (w_2 a)^{j+1}w_r \in L$ since $L$ is aperiodic, see
  Lemma~\ref{lem:lsla-aperiodic}, and $j \geq |A_L|$. This can be rewritten as
  \[w_\ell w_1 ( aw_1 )^{j-1} a w_m w_2 ( aw_2)^{j-1} (aw_2a w_r) \in L\;.\] As
  $L$ is closed under \lsla, and $n < j$, this implies
  \[w_\ell w_1 ( aw_1 )^{j-1} ( aw_2)^{j-1} (aw_2a w_r) \in L\;.\] This can be
  rewritten into $w_\ell (w_1 a)^{j} (w_2a)^{j} w_r \in L$.

  (4) $\Rightarrow$ (2): Let $L$ be closed under \rsla. We will prove that $A_L$
  satisfies the \rscp, that is, if there are two states $q_1, q_2$ in $A_L$ with
  $q_1 \leadsto q_2$ and $w_1 \in \loopwords(q_1)$, $w_2 \in \loopwords(q_2)$,
  such that $w_1$ and $w_2$ end with the same letter, then $(w_2 a)^N \cL_{q_2}
  \subseteq \cL_{q_1}$. Let $q_1$, $q_2$ be such states. Then there exist $w_\ell,
  w_m$ with $q_1 =\delta_L(i_L, w_\ell)$ and $q_2 = \delta_L(q_1,w_m)$. If
  $\cL_{q_2} = \emptyset$, we are done. So let us assume there is a word $w_r \in
  \cL_{q_2}$. We define $w'_r = w_2^N w_r$. Due to construction, we have $w_\ell
  w_1^* w_m w_2^* w'_r \subseteq L$. Since $L$ is closed under \rsla, there is
  an $i \in \nat$ such that $w_\ell w_1^i w_2^i w'_r \in L$. Since we have a
  deterministic automaton and $q_1 =\delta_L(i_L, w_\ell w_1^i)$ this implies
  that $w_2^i w'_r = w_2^i w_2^N w_r \in \cL_{q_1}$. We now use that $L$ is
  aperiodic due to Lemma~\ref{lem:rsla-aperiodic} to infer that $w_2^N w_r \in
  \cL_{q_1}$.

  (2) $\Rightarrow$ (1):  Let $A_L$ satisfy the \rscp. We will show that if there exist states $q_1,q_2 \in
  Q_L$ and words $w_1, w_2 \in \Sigma^*$ with $a w_1 \in \loopwords(q_1)$ and $a
  w_2 \in \loopwords(q_2)$ and $q_1 \leadsto q_2$, then $(aw_2)^N \cL_{q_2}
  \subseteq \cL_{q_1}$. Let $q_1, q_2$ be such states and $w_1, w_2$ as above. We
  define $q'_1 = \delta_L(q_1, w_1)$ and $q'_2 = \delta_L(q_2, w_2)$. Since
  $A_L$ is deterministic, the construction implies that $w_1 a \in
  \loopwords(q'_1)$ and $w_2 a \in \loopwords(q'_2)$. Furthermore, it holds that
  (i) $\cL_{q'_1}= a^{-1} \cL_{q_1}$ and (ii) $w_2 \cL_{q_2} \subseteq \cL_{q'_2}$. With this we
  will show that $(w_2 a)^N \cL_{q'_2} \subseteq \cL_{q'_1}$ implies $(a w_2)^N \cL_{q_2}
  \subseteq \cL_{q_1}$. Let $(w_2 a)^N \cL_{q'_2} \subseteq \cL_{q'_1}$. Adding an $a$ left
  hand, yields $(a w_2)^N a \cL_{q'_2} \subseteq a \cL_{q'_1} \subseteq \cL_{q_1}$ because
  of (i). We use (ii) to replace $\cL_{q'_2}$ to get: $(a w_2)^{N+1} \cL_{q_2} \subseteq
  \cL_{q_1}$. Since $L$ is aperiodic, see Lemma~\ref{lem:rscp-aperiodic}, this
  is equivalent to $(a w_2)^{N} \cL_{q_2} \subseteq \cL_{q_1}$.
\end{proof}
  \begin{cor}\label{cor:leftsync-abbrev}
    If a regular language $L$ satisfies Definition~\ref{def:leftsyncpower} and
    $N = |A_L|$ then, for all $i > N^2+N$ and for all words $w_\ell, w_m, w_r \in
    \Sigma^*$ and all words $w_1 = a w_1'$ and $w_2 = a w_2'$ (resp., $w_1 =
    w_1' a$ and $w_2 = w_2' a$) we have that $w_\ell w_1^i w_m w_2^i w_r \in L$
    implies $w_\ell w_1^i w_2^i w_r \in L$.
  \end{cor}
  \begin{proof}
    This immediately follows from the proof of (1) $\Rightarrow$ (3).
  \end{proof}

In
Theorem~\ref{theo:trichotomy} we will show that, if NL $\neq$ NP, the languages
$L$ that satisfy the above properties are precisely those for which $\rtq(L)$ is tractable. To
simplify terminology, we will henceforth refer to this class as \ttract.
\begin{defi}\label{def:ttract}
  A regular language $L$ belongs to \ttract if $L$ satisfies one of
  the equivalent conditions in Theorem~\ref{theo:equivalence}.
\end{defi}
For example, $(ab)^*$ and $(abc)^*$ are in \ttract, whereas $a^*ba^*$, $(aa)^*$
and $(aba)^*$ are not. The following property immediately follows from the
definition of \ttract.
\begin{obs}\label{obs:ttract-to-aperiodic}
  Every regular expression for which each alphabet symbol under a Kleene star
  occurs at most once in the expression defines a language in \etract.
\end{obs}
A special case of these expressions are those in which every alphabet symbol
occurs at most once. These are known as \emph{single-occurrence regular
  expressions (SORE)}~\cite{BexNSV-tods10}. SOREs were studied in the context of
learning schema languages for XML~\cite{BexNSV-tods10}, since they occur very
often in practical schema languages.

\subsection{The inner Structure of minimal DFAs in \texorpdfstring{\ttract}{Ttract}}
The components of minimal DFAs of languages in \ttract have a very special form.
The insights provided in this section are used in Section~\ref{sec:trichotomy}
to show trichotomy results for \ttract, and in
Section~\ref{sec:syntactic-characterization} to give a syntactic
characterization of languages in \ttract.

\begin{lem}\label{lem:component}
	Let $L \in \ttract$, $a \in \Sigma$, $C$ be a component of $A_L$, and $q_1, q_2 \in C$.
	If there exist $w_1a \in \loopwords(q_1)$ and $w_2a \in \loopwords(q_2)$,
	then, for all $\sigma \in \Sigma$, we have that $\delta_L(q_1,\sigma)\in C$ if
	and only if $\delta_L(q_2,\sigma)\in C$.
\end{lem}
\begin{proof} Let $q_1\neq q_2$ be two states in $C$. Let
	$\sigma$ satisfy $\delta_L(q_1,\sigma)\in C$ and let $w\in \loopwords(q_1)\cap
	\sigma \Sigma^* a$. Such a $w$ exists since $\delta_L(q_1,\sigma)\in C$
	and $\delta_L(q_1,w_1a)=q_1$.
	Let $q_3 = \delta_L(q_2,w^N)$. We will prove that $q_1=q_3$, which
	implies that $\delta_L(q_2,\sigma)\in C$. As $L$ is aperiodic, $w \in
	\loopwords(q_3)$.
	Consequently, there
	is an $n \in \nat$ such that $w^n \cL_{q_3} \subseteq \cL_{q_1}$ by
	Definition~\ref{def:leftsync}. Since $w\in \loopwords(q_1)$, this also implies
	$\cL_{q_3}\subseteq \cL_{q_1}$. Furthermore, $q_2$ has a loop ending with $a$ and
	$A_L$ satisfies the \rscp, so $w^N\cL_{q_1} \subseteq \cL_{q_2}$ by Corollary~\ref{cor:leftsync}. Hence,
	$\cL_{q_1}\subseteq (w^N)^{-1}\cL_{q_2}$ and, by definition of $q_3$, we have
	$(w^N)^{-1}\cL_{q_2} = \cL_{q_3}$. So we showed $\cL_{q_3} \subseteq \cL_{q_1}$ and $\cL_{q_1}
	\subseteq \cL_{q_3}$ which, by minimality of $A_L$, implies $q_1 = q_3$.
\end{proof}

The following is a direct consequence thereof.
\begin{cor}\label{cor:component}
	Let $L \in \ttract$, $a \in \Sigma$, $C$ be a component of $A_L$, and $q_1, q_2\in C$. If there
	exist $w_1 a \in \loopwords(q_1)$ and $w_2 a \in \loopwords(q_2)$, then
	$\delta_L(q_1,w) \in C$ if and only if $\delta_L(q_2,w)\in C$ for all words $w
	\in \Sigma^*$.
\end{cor}

\begin{lem}\label{lem:sameWord}
	Let $A_L$ satisfy the \lscp. If states $q_1$ and $q_2$ belong to the same
	component of $A_L$ and $\loopwords(q_1) \cap \loopwords(q_2) \neq
	\emptyset$, then $q_1 = q_2$.
\end{lem}
\begin{proof}
	Let $q_1,q_2$ be as stated and let $w$ be a word in $\loopwords(q_1)\cap
	\loopwords(q_2)$. According to Definition~\ref{def:leftsync}, there exists an
	$n \in \nat$ such that $w^n
	\cL_{q_2}\subseteq \cL_{q_1}$. Since $w \in \loopwords(q_1)$, this implies that $\cL_{q_2}\subseteq
	\cL_{q_1}$. By symmetry, we have $\cL_{q_2}=\cL_{q_1}$, which implies $q_1 = q_2$, since
	$A_L$ is the minimal DFA\@.
\end{proof}

To this end, we obtain the following synchronization property for $A_L$.
\begin{lem}\label{lem:sameState}
	Let $L\in \ttract$, let $C$ be a component of $A_L$, let $q_1, q_2\in C$, and
	let $w$ be a word of length
	$N^2$. If $\delta_L(q_1,w) \in C$ and $\delta_L(q_2,w) \in C$, then
	$\delta_L(q_1,w)=\delta_L(q_2,w)$.
\end{lem}
\begin{proof}
	Assume that $w=a_1 \cdots a_{N^2}$. For each $i$ from $0$ to $N^2$ and $\alpha
	\in \{1,2\}$, let $q_{\alpha,i}=\delta_L(q_\alpha,a_1\cdots a_i)$. Since there
	are at most $N^2$ distinct pairs $(q_{1,i}, q_{2,i})$, there exist $i,j$ with
	$0\leq i<j \leq N^2$ such that $q_{1,i}=q_{1,j}$ and $q_{2,i}=q_{2,j}$. Since
	$\delta_L(q_1,w) \in C$ and $\delta_L(q_2,w) \in C$, $q_{1,i},q_{2,i} \in C$.
	Let $w' = a_{i+1} \cdots a_j$. We have $w' \in \loopwords(q_{1,i}) \cap
	\loopwords(q_{2,i})$, hence $q_{1,i}=q_{2,i}$ by Lemma~\ref{lem:sameWord}. As
	a consequence, $\delta_L(q_1,w)=\delta_L(q_2,w)$.
\end{proof}

Furthermore, we show that every language in \ttract satisfies an inclusion property which is stronger than indicated by Definition~\ref{def:leftsync}. That is, we show that it is not necessary to repeat some word $w_2$ multiple times.  Instead, we show that any word $w$ that stays in a component, given that $w$ is long enough and starts with a suitable symbol, already implies an inclusion property.
\begin{lem}\label{lem:Bagan9}
	Let $L\in \ttract$, $a \in \Sigma$ and let $q_1,q_2$ be two states such that $q_1 \leadsto
	q_2$ and $\loopwords(q_1)\cap a\Sigma^* \neq
	\emptyset$.
	Let $C$ be
	the component of $A_L$ that contains $q_2$. Then, \[\cL_{q_2} \cap
	L_{q_2}^{a}\Sigma^* \subseteq \cL_{q_1}\] where $L_{q_2}^{a}$ is the set
	of words $w$ of length $N^2$ that start with $a$ and such that
	$\delta_L(q_2,w)\in C$.
\end{lem}
\begin{proof}
	If $\loopwords(q_2) = \emptyset$, then $\cL_{q_2} \cap
	L_{q_2}^{a}\Sigma^* = \emptyset$ and the inclusion trivially holds. Therefore
	we assume from now on that $\loopwords(q_2) \neq \emptyset$.
	Since the proof of this lemma requires a number of different states and words, we
	provide a sketch in Figure~\ref{fig:Bagan9}.
	Let $w \in \cL_{q_2}\cap L_{q_2}^{a}\Sigma^*$, $u$ be the prefix of
	$w$ of length $N^2$ and $w'$ be the suffix of $w$ such that $w=uw'$.
	Since $q_2$ and $\delta_L(q_2,u)$ are both in the same \component $C$, there exists a word $v$ with $uv \in \loopwords(q_2)$.
	Corollary~\ref{cor:leftsync} implies that
	\begin{equation}\label{eq:lem:Bagan9:1}
		(uv)^N\cL_{q_2} \subseteq \cL_{q_1}\;.
	\end{equation}
	Let $q_3 = \delta_L(q_1,(uv)^N)$. Due to aperiodicity we have $uv \in
	\loopwords(q_3)$. Since $A_L$ is deterministic, this implies $\cL_{q_3} =
	((uv)^N)^{-1} \cL_{q_1}$ and, together with Equation~\eqref{eq:lem:Bagan9:1} that
	\begin{equation}\label{eq:lem:Bagan9:2}
		\cL_{q_2} \subseteq \cL_{q_3}\;.
	\end{equation}
	We now show that there is a prefix $u_1$ of $u$ such that $\delta_L(q_1,u_1) =
	q$ and $\delta_L(q_3,u_1)=q'$ with $\loopwords(q) \cap \loopwords(q') \neq
	\emptyset$. %
	Assume that $u=a_1 \cdots a_{N^2}$. Let $q_{\alpha,0}=q_\alpha$ and, for each
	$i$ from $1$ to $N^2$ and $\alpha \in \{1,3\}$, let
	$q_{\alpha,i}=\delta_L(q_\alpha,a_1\cdots a_i)$. Since there are at most $N^2$
	distinct pairs $(q_{1,i}, q_{3,i})$, there exist $i,j$ with $0\leq i<j \leq
	N^2$ such that $q_{1,i}=q_{1,j}$ and $q_{3,i}=q_{3,j}$. Let $u_1 = a_1 \cdots
	a_i$ and $u_2 = a_{i+1} \cdots a_j$. We have $u_2 \in \loopwords(q_{1,i}) \cap
	\loopwords(q_{3,i})$. We define $q = \delta_L(q_1,u_1)$ and
	$q' = \delta_L(q_3,u_1)$. Since $q \leadsto q'$ and $u_2 \in
	\loopwords(q) \cap \loopwords(q')$, Corollary~\ref{cor:leftsync} implies
	$u_2^N \cL_{q'}\subseteq \cL_{q}$. Since $u_2 \in \loopwords(q)$, we also have that
	\begin{equation}\label{eq:lem:Bagan9:3}
		\cL_{q'}\subseteq \cL_{q}\;.
	\end{equation}
	By definition of $q$ and the determinism of $A_L$, we have that $\cL_{q} = u_1^{-1}
	\cL_{q_1}$. Thus, Equation~\eqref{eq:lem:Bagan9:3} implies $\cL_{q'} \subseteq u_1^{-1}
	\cL_{q_1}$. The definition of $q'$ implies that $\cL_{q'} = u_1^{-1} \cL_{q_3}$, so
	$u_1^{-1} \cL_{q_3} \subseteq u_1^{-1} \cL_{q_1}$.
	In other words, we have $\cL_{q_3}\cap u_1 \Sigma^* \subseteq \cL_{q_1}\cap u_1
	\Sigma^* $. Since $u_1$ is a prefix of $u$, and by Equation~\eqref{eq:lem:Bagan9:2}, we also have
	$\cL_{q_2}\cap u \Sigma^* \subseteq \cL_{q_1}$. This implies that $w \in \cL_{q_1}$, which concludes the proof.
\end{proof}

\begin{figure}[t] \centering
	\begin{tikzpicture}[->,>=latex,shorten >=.5pt,auto,node distance=1.3cm,
		inner sep = 0.7mm, initial text= $\cdots$]

		\node[state,initial,minimum size=17pt] (q1) at (-3,0) {$q_1$};
		\node[state,minimum size=17pt] (q2) at (-1,1) {$q_2$};
		\node[state,minimum size=17pt] (q3) at (.7,0) {$q_3$};
		\node[state,minimum size=17pt] (q4) at (-1.5,0) {$q$};
		\node[state,minimum size=17pt] (q5) at (3.5,0) {$q'$};
		\node[state,minimum size=0pt] (q6) at (1,1) {};
		\node[state,minimum size=0pt, accepting] (q7) at (3,1.5) {};

		\path (q1) edge  [->, loop above] node {$a..$} (q1);
		\path (q2) edge [->, loop above] node {$a..$} (q2);

		\path[draw,<-,decorate,
		decoration={snake,amplitude=.3mm,segment length=2.5mm,pre=lineto,pre
			length=5pt}] (q2) -- (q1);

		\path (q4) edge [->,swap] node {$u_1^{-1}(uv)^N$} (q3); \path (q3) edge [->,
		bend left = 20] node {$u_1$} (q5); \path (q5) edge [->, bend left = 20] node
		{$u_1^{-1}uv$} (q3);

		\path (q4) edge [->, loop below] node {$u_2$} (q4); \path (q1) edge
		[->,swap] node {$u_1$} (q4); %
		\path (q5) edge [->, loop right] node {$u_2$} (q5);

		\path (q2) edge [->, bend left = 20] node {$u$} (q6);
		\path (q6) edge [->,bend left =20] node {$v$} (q2);
		\path (q6) edge [->, out=20, in = -130] node {$w'$} (q7);
	\end{tikzpicture}
	\caption{Sketch of the proof of Lemma~\ref{lem:Bagan9}}%
	\label{fig:Bagan9}
\end{figure}

\subsection{A Syntactic Characterization}\label{sec:syntactic-characterization}
The goal of this section is to give a better understanding of languages in
\ttract. We provide a syntactic definition, which will allow to construct
languages in \ttract. More precisely, we will show that every language of a
``memoryless component'' is in \ttract. And if memoryless components are
connected with ``consistent jumps'', then the language is again in \ttract. We
show that all languages in \ttract can be constructed in this way. Using this
modular principle, systems with graphical search queries could enable users to
``click'' a language in \ttract together. Note that this section is quite
technical and detached from the rest of the article, thus it can be skipped.

As we have seen before, regular
expressions in which every symbol occurs at most once define
languages in \etract. We will define a similar notion on automata.
\begin{defi}\label{def:SO} A component $C$ of some NFA $A$ is
  called \emph{memoryless},
  if for each symbol
  $a \in \Sigma$, there is at most one state $q$ in $C$, such that there is a transition $(p,a,q)$ with $p$ in $C$.
\end{defi}

In this section, we will prove the following theorem which provides (in a non-trivial proof that requires several
steps) a syntactic condition for languages in \ttract. The syntactic condition
is item (4) of the theorem, which we define after its statement. Condition (5)
imposes an additional restriction on condition (4).
\begin{thm}\label{thm:ttractNFA}
	For a regular language $L$, the following properties are equivalent:
	\begin{enumerate}[(1)]
		\item $L \in \ttract$\label{thm:ttractNFA:ttract}
		\item There exists an NFA $A$ for $L$ that satisfies the \lscp.
		\item There exists an NFA $A$ for $L$ that satisfies the \lscp and only has
		memoryless components.\label{thm:ttractNFA:memorylessNFA}
		\item There exists a detainment automaton for $L$ with consistent jumps.
		\item There exists a detainment automaton for $L$ with consistent jumps
		and only memoryless components.\label{thm:ttractNFA:memorylessCNFA}
	\end{enumerate}
\end{thm}

\noindent
To define detainment automata, we use \emph{finite automata with counters or CNFAs} from Gelade et al.~\cite{GeladeGM-sicomp12}, which we slightly adapt to make the construction
easier.\footnote{The adaptation is that we let counters decrease instead of
	increase. Furthermore, it only needs zero-tests.}

\newcommand{\nfac}{\text{CNFA}\xspace}

We recall the definition of counter NFAs from Gelade et
al.~\cite{GeladeGM-sicomp12}. We introduce a minor difference, namely that
counters count down instead of up, since this makes our construction easier to
describe. Furthermore, since our construction only requires a single counter,
zero tests, and setting the counter to a certain value, we immediately
simplify the definition to take this into account.

Let $c$ be a \emph{counter variable}, taking values in $\nat$. A \emph{guard}
on $c$ is a statement $\gamma$ of the form $\true$ or $c = 0$. We denote by $c
\models \gamma$ that $c$ satisfies the guard $\gamma$. In the case where
$\gamma$ is \true, this is trivially fulfilled and, in the case where $\gamma$
is $c = 0$, this is fulfilled if $c$ equals $0$. By $G$ we denote the set of
guards on $c$.
An \emph{update} on $c$ is a statement of the form $c:=c-1$, $c:=c$, or $c:=k$ for some
constant $k \in \nat$. By $U$ we denote the set of updates on $c$.

\begin{defi}\label{def:counterautomaton}
	A \emph{nondeterministic counter automaton} (\nfac) with a single counter is a 6-tuple $A =
	(Q,I,c,\delta,F,\tau)$ where $Q$ is the finite set of states; $I \subseteq  Q$
	is a set of initial states; $c$ is a counter variable; $\delta
	\subseteq Q \times \Sigma \times G \times Q \times U$ is
	the transition relation; and $F \subseteq Q$ is the set of accepting states.
	Furthermore, $\tau\in\nat$ is a constant such that every update of the
	form $c:=k$ has $k\leq \tau$.
\end{defi}

Intuitively, $A$ can make a transition $(q,a,\gamma;q',\pi)$ whenever it is in
state $q$, reads $a$, and $c \models \gamma$, i.e., guard $\gamma$ is true
under the current value of $c$. It then updates $c$ according to the update
$\pi$, in a way we explain next, and moves into state $q'$. To explain the
update mechanism formally, we introduce the notion of configuration. A
\emph{configuration} is a pair $(q,\ell)$ where $q \in Q$ is the current state
and $\ell \in \nat$ is the value of $c$. Finally, an update $\pi$ defines
a function $\pi: \nat \to \nat$ as follows. If $\pi = (c:=k)$ then
$\pi(\ell) = k$ for every $\ell \in \nat$. If $\pi = (c:=c-1)$ then $\pi(\ell)
= \max(\ell-1,0)$. Otherwise, i.e., if $\pi = (c:=c)$, then $\pi(\ell) =
\ell$. So, counters never become negative.

An \emph{initial configuration}  is $(q_0, 0)$ with $q_0 \in I$. A configuration
$(q,\ell)$ is \emph{accepting} if $q \in F$ and $\ell=0$.
A configuration
$\alpha'=(q',\ell')$ \emph{immediately follows} a configuration
$\alpha=(q,\ell)$ by reading $a \in \Sigma$, denoted $\alpha\rightarrow_{a}
\alpha'$, if there exists $(q,a,\gamma;q',\pi) \in \delta$ with $c \models
\gamma$ and $\ell' = \pi(\ell)$.

For a string $w = a_1\cdots a_n$ and two configurations $\alpha$ and
$\alpha'$, we denote by $\alpha \Rightarrow_{w} \alpha'$ that $\alpha
\rightarrow_{a_1} \cdots \rightarrow_{a_n} \alpha'$. A configuration $\alpha$
is \emph{reachable} if there exists a string $w$ such that $\alpha_0
\Rightarrow_w \alpha$ for some initial configuration $\alpha_0$. A string $w$ is \emph{accepted} by $A$ if
$\alpha_0\Rightarrow_{w}\alpha_f$ where $\alpha_0$ is an initial configuration and $\alpha_f$ is an accepting configuration.
We denote by $L(A)$ the set of strings accepted by $A$.

It is easy to see that \nfac accept precisely the regular languages. (Due to the value
$\tau$, counters are always bounded by a constant.)

Let $A$ be a CNFA with one counter $c$. Initially, the
counter has value 0. The automaton has transitions of the form $(q_1,a,P;q_2,U)$
where $P$ is a precondition on $c$ and $U$ an update operation on $c$. For
instance, the transition $(q_1,a,c=5;q_2,c:=c-1)$ means: if $A$ is in state
$q_1$, reads $a$, and the value of $c$ is five, then it can move to $q_2$ and
decrease $c$ by one. If we decrease a counter with value zero, its value remains
zero. We denote the precondition that is always fulfilled by $\true$.

We say that $A$ is a \emph{detainment automaton} if, for every component $C$ of $A$:
\begin{itemize}
	\item every transition inside $C$ is of the form $(q_1,a,\true;q_2,c:=c-1)$;
	\item every transition that leaves $C$ is of the form $(q_1,a,c=0;q_2,c:=k)$ for
	some $k \in \nat$;\footnote{If $q_2$ is in a trivial component, then $k$ should be $0$ for the transition to be useful.}
\end{itemize}
Intuitively, if a detainment automaton enters a non-trivial component
$C$, then it must stay there for at least
some number of steps, depending on the value of the counter $c$. The counter $c$
is decreased for every transition inside $C$ and the automaton can only leave
$C$ once $c = 0$. We say that $A$ has \emph{consistent jumps}  if, for every pair of components $C_1$ and $C_2$, if
$C_1 \leadsto C_2$ and there are transitions $(p_i,a,\true; q_i,c:=c-1)$ inside
$C_i$ for all $i \in \{1,2\}$, then there is also a transition $(p_1,a, P; q_2,U)$ for some $P \in
\{\true, c=0\}$ and some update $U$.\footnote{The values of $P$ and $U$ depend
	on whether $C_1$ is the same as $C_2$ or not.} We illustrate this in Figure~\ref{fig:consistent-jumps}.
We note that $C_1$ and $C_2$ may be the same component.  The consistent jump
property is the syntactical counterpart of the left-synchronized containment
property. The \emph{memoryless} condition carries over naturally to CNFAs,
ignoring the counter.

\begin{figure}[t] \centering
	\begin{minipage}{.3\linewidth}
		\begin{tikzpicture}[auto,>=latex]
			\node [label={[xshift=0.05cm, yshift=-.3cm]$C_1$}] (c1) at (0.25,1) {};
			\node [label={[xshift=0.05cm, yshift=-.3cm]$C_2$}] (c2) at (2.75,1) {};
			\node [label={[xshift=-0.25cm, yshift=-.35cm]$p_1$}] (q1) at (0,0) {};
			\node  [label={[xshift=0.25cm, yshift=-.35cm]$q_1$}] (q2) at (.5,.5) {};
			\node [label={[xshift=-0.25cm, yshift=-.35cm]$p_2$}] (q3) at (2.5,0) {};
			\node [label={[xshift=0.25cm, yshift=-.35cm]$q_2$}] (q4) at (3,0.5) {};

			\foreach \i in {1,2,3,4}{
				\fill (q\i) circle (2pt);
			}

			\node(ell1) at ($(q1)!0.5!(q2)$) {};
			\draw (ell1) ellipse (0.8cm and .65cm);

			\node(ell2) at ($(q3)!0.5!(q4)$) {};
			\draw (ell2) ellipse (0.8cm and .65cm);

			\path[->]
			(q1) edge node  {} (q2)
			(q3) edge node {} (q4);

			\node [label={[xshift=-0.2cm, yshift=-.22cm]$a$}] (v3) at ($(q1)!0.5!(q2)$) {};
			\node [label={[xshift=-0.2cm, yshift=-.22cm]$a$}] (v4) at ($(q3)!0.5!(q4)$) {};
		\end{tikzpicture}
	\end{minipage}
	\hspace{.1cm}
	$\Rightarrow$
	\hspace{.1cm}
	\begin{minipage}{.3\linewidth}
		\begin{tikzpicture}[auto,>=latex]
			\node [label={[xshift=0.05cm, yshift=-.3cm]$C_1$}] (c1) at (0.25,1) {};
			\node [label={[xshift=0.05cm, yshift=-.3cm]$C_2$}] (c2) at (2.75,1) {};
			\node [label={[xshift=-0.25cm, yshift=-.35cm]$p_1$}] (q1) at (0,0) {};
			\node  [label={[xshift=0.25cm, yshift=-.35cm]$q_1$}] (q2) at (.5,.5) {};
			\node [label={[xshift=-0.25cm, yshift=-.35cm]$p_2$}] (q3) at (2.5,0) {};
			\node [label={[xshift=0.25cm, yshift=-.35cm]$q_2$}] (q4) at (3,0.5) {};

			\foreach \i in {1,2,3,4}{
				\fill (q\i) circle (2pt);
			}

			\node(ell1) at ($(q1)!0.5!(q2)$) {};
			\draw (ell1) ellipse (0.8cm and .65cm);

			\node(ell2) at ($(q3)!0.5!(q4)$) {};
			\draw (ell2) ellipse (0.8cm and .65cm);

			\path[->]
			(q1) edge node  {} (q2)
			(q3) edge node {} (q4);

			\node [label={[xshift=-0.2cm, yshift=-.22cm]$a$}] (v3) at ($(q1)!0.5!(q2)$) {};
			\node [label={[xshift=-0.2cm, yshift=-.22cm]$a$}] (v4) at ($(q3)!0.5!(q4)$) {};

			\path(q4) edge [ in =-20, out = 140,draw,<-,decorate,
			decoration={snake,amplitude=.0mm,segment length=2.5mm,pre=lineto,pre
				length=5pt}]  (q1);
			\node [label={[xshift=0.25cm, yshift=-.2cm]$a$}] (q5) at ($(q1)!0.5!(q3)$) {};

		\end{tikzpicture}
	\end{minipage}
	\caption{Consistent jump condition (simplified, i.e.:\@ without preconditions, counter and update) used in Theorem~\ref{theo:classification}. $C_1$ and $C_2$ are components (not necessarily different) such that $C_2$ is reachable from $C_1$.}%
	\label{fig:consistent-jumps}
\end{figure}

	\begin{proof}[Proof sketch of Theorem~\ref{thm:ttractNFA}]
		The implications (3) $\Rightarrow$ (2) and (5) $\Rightarrow$ (4) are
		trivial. We sketch the proofs of (1) $\Rightarrow$ (5) $\Rightarrow$ (3) and (4)
		$\Rightarrow$ (2) $\Rightarrow$ (1) below, establishing the theorem.

		(1) $\Rightarrow$ (5) uses a very technical construction that
		essentially exploits that---if the automaton stays in the same
		component for a long time---the reached state only depends on
		the last $N^2$ symbols read in the component. This is formalized in
		Lemma~\ref{lem:sameState} and allows us to merge any pair of two
		states $p,q$ which contradict that some component is memoryless. To
		preserve the language, words that stay in some component $C$ for
		less than $N^2$ symbols have to be dealt with separately,
		essentially avoiding the component altogether. Finally, the \lscp
		allows us to simply add transitions required to satisfy the
		consistent jumps property without changing the language.

		(5) $\Rightarrow$ (3) and (4) $\Rightarrow$ (2): We convert a given
		CNFA to an NFA by simulating the counter (which is bounded) in the
		set of states.  The consistent jump property implies the \lscp on
		the resulting NFA\@. The property that all components are memoryless
		is preserved by the construction.

		(2) $\Rightarrow$ (1): One can show that the \lscp is invariant
		under the powerset construction.
		\phantom\qedhere %
	\end{proof}

  The following lemma is the implication (1) $\Rightarrow$ (5) from Theorem~\ref{thm:ttractNFA}
  \begin{lem}\label{lem:memoryless}
    If  $L \in \ttract$, then there exists a detainment automaton for $L$ with consistent jumps
    and only memoryless components.
  \end{lem}
  \begin{proof}
    Let $A_L=(Q_L,\Sigma,i_L,F_L,\delta_L)$ be the minimal DFA for
    $L$.  The proof goes as follows: First, we define a CNFA $A$ with
    two counters.  Second, we show that we can convert $A$ to an
    equivalent CNFA $A'$ with only one counter that is a detainment
    automaton with consistent jumps and only memoryless
    components. This conversion is done by simulating one of the
    counters using a bigger set of states. Last, we show that
    $L(A)=L(A_L)$, which shows the lemma statement as $L(A)=L(A')$.

    Before we start we need some additional notation.  We write
    $p_1 \curvearrowright^a q_2$ to denote that
    $C(p_1) \leadsto C(q_2)$ and there are states $q_1 \in C(p_1)$ and
    $p_2 \in C(q_2)$ such that $(p_i,a,q_i) \in \delta_L$ for
    $i \in \{1,2\}$.
    Let $q$ be a state,
    then we write $\LoopSym{q}$ to denote the set of symbols $a$, such
    that there is a word $w=aw' \in \loopwords(q)$.

    Let ${\sim} \subseteq Q_L \times Q_L$ be the smallest equivalence
    relation over $Q_L$ that satisfies $p \sim q$ if $C(p)=C(q)$ and
    $\LoopSym{p} \cap \LoopSym{q} \neq \emptyset$.  For $q \in Q_L$,
    we denote by $[q]$ the equivalence class of $q$. By $[Q_L]$ we
    denote the set of all equivalence classes. We also write $[C]$ to
    denote the equivalence classes that only use states from some
    component $C$. We extend the notion $C(q)$ to $[Q_L]$, i.e.,
    $C([q])=C(q)$ for all $q \in Q_L$.

    We will use the following observation that easily follows from
    Lemma~\ref{lem:component} using the definition of $\sim$.
    \begin{obs}\label{obs:component}
      Let $q_1,q_2$ be states with $[q_1]=[q_2]$, then for all
      $a \in \Sigma$ it holds that $\delta_L(q_1,a) \in C(q_1)$ if and only if
      $\delta_L(q_2,a) \in C(q_1)$.
    \end{obs}

    We define a CNFA $A=(Q,I,c,d,\delta,F,N^2)$ that has two
    counters $c$ and $d$. The counter $c$ is allowed to have any
    initial value from $[0,N^2]$, while the counter $d$ has initial
    value $0$. We note that we will eliminate counter $c$ when
    converting to a one counter automaton, thus this is not a
    contradiction to the definition of CNFA with one counter that we
    use.

    We use $Q' = Q_L \cup [Q_L]$, i.e., we can use the states from
    $A_L$ and the equivalence classes of the equivalence relation
    $\sim$. The latter will be used to ensure that \components are memoryless,
    while the former will only be used in
    trivial \components. We use $I=\{i_L,[i_L]\}$
    and $F=F_L$.

  \newcommand{\alignDelta}[1]{\mathrlap{#1}{\phantom{\delta_x}}}
  \begin{align*}
    \alignDelta{\delta_\circlearrowright^1} \; =& \;\; \{\;(q_1,a,\{c>0,d=0\};q_2,\{c := c-1\}) \;\mid\; (q_1,a,q_2) \in \delta_L, C(q_1)=C(q_2) \;\}  \\
    \alignDelta{\delta_\circlearrowright^2} \; =& \;\; \{\;([q_1],a,\{c=N^2\};[q_2],\{d:=d-1\}) \;\mid\; (q_1,a,q_2) \in \delta_L, C(q_1)=C(q_2) \;\}  \\
    \alignDelta{\delta_\circlearrowright^3} \; =& \;\; \{\;([q_1],a,\{c=N^2,d=0\};q_2,\{c:=c-1\}) \;\mid\; (q_1,a,q_2) \in \delta_L, C(q_1)=C(q_2) \;\} \\
    \alignDelta{\delta_\rightarrow^1} \; =& \;\; \{\;(q_1,a,\{c=0,d=0\};q_2,\{c:=i\}) \mid \; (q_1,a,q_2) \in \delta_L, C(q_1)\neq C(q_2), i \in [0,N^2-1] \;\} \\
    \alignDelta{\delta_\rightarrow^2} \; =& \;\; \{\;(q_1,a,\{c=0,d=0\};[q_2],\{c:=N^2\}) \mid \; (q_1,a,q_2) \in \delta_L, C(q_1)\neq C(q_2) \;\} \\
    \alignDelta{\delta_\curvearrowright} \; =& \;\; \{ \; ([q_1],a,\{c=N^2,d=0\};[q_2],\{d:=N^2\}) \;\mid\; q_1 \curvearrowright^a q_2, C(q_1) \neq C(q_2) \;\}\\
    \alignDelta{\delta} \; =& \;\; \delta_\circlearrowright^1 \;\;\cup\;\; \delta_\circlearrowright^2 \;\;\cup\;\; \delta_\circlearrowright^3 \;\;\cup\;\; \delta_\rightarrow^1 \;\;\cup\;\; \delta_\rightarrow^2 \;\;\cup\;\; \delta_\curvearrowright%
  \end{align*}

  We say that a component $C$ of $A_L$ is a \emph{long run component}
  of a given word $w=a_1 \cdots a_n$, if
  $|\{ i \mid \delta(i_L, a_1\cdots a_i) \in C\}|>N^2$, i.e., if the
  run stays in $C$ for more than $N^2$ symbols.
  All other components
  are \emph{short run components}.

  For short run components, we use states from $Q_L$. We use the
  counter $c$ to enforce that these parts are indeed short. For long
  run components, we first use states in $[Q_L]$. Only the last $N^2$
  symbols in the component are read using states from $Q_L$. The \lscp
  guarantees that for long run components the precise state is not
  important, which allows us to make these components memoryless.

  The transition relation is divided into transitions  between states from the same
  component of $A_L$ (indicated by
  $\delta_\circlearrowright=\delta_\circlearrowright^1
  \cup\delta_\circlearrowright^2\cup\delta_\circlearrowright^3$) and
  transitions between different components (indicated by
  $\delta_\rightarrow=\delta_\rightarrow^1\cup\delta_\rightarrow^2$).
  Transitions in $\delta_\curvearrowright$ are added to satisfy the
  consistent jumps property. They are the only transitions that
  increase the counter $d$. This is necessary, as the \lscp only talks
  about the language of the state reached after staying in the
  component for some number of symbols. If we added the
  transitions in $\delta_\curvearrowright$ without using the
  counter, we would possibly add additional words to the language.
  This concludes the definition of $A$.

  We now argue that the automaton
  $A'=(Q'\times [0,N^2],i_L,d,\delta',F\times\{0\},N^2)$ derived from $A$ by
  pushing the counter $c$ into the states is a detainment automaton
  with consistent jumps that only has memoryless components.  The
  states of $A'$ have two components, first the state of $A$ and
  second the value of the second counter that is bounded by $N^2$. We
  do not formally define $\delta'$. It is derived from $\delta$ in the
  obvious way, i.e., by doing the precondition checks that depend on
  $c$ on the second component of the state. Similarly, updates of $c$
  are done on the second component of the states.

  It is straightforward to see that $A'$ is a detainment automaton
  with consistent jumps that only has memoryless components using the
  following observations:
  \begin{itemize}
  \item Every transition in $A$ that does not have $c=N^2$ before and
    after the transition requires $d=0$.
  \item Let $\Cuts$ be the set of  \components of $A$, then the set of \components of $A'$ is $\{[C] \times \{N^2\} \mid C \in \Cuts\}$.
  \end{itemize}
  The consistent jumps are guaranteed by the transitions in
  $\delta_\curvearrowright$. As $A'$ only has memoryless components,
  the consistent jump property is trivially satisfied for states inside
  the same component.

  \newcommand{\jump}{\mathsf{countdown}}

  We now show that $L(A_L) \subseteq L(A)$. Let $w=a_1\cdots a_n$ be
  some string in $L(A_L)$ and $q_0 \rightarrow \cdots \rightarrow q_n$
  be the run of $A_L$ on $w$.
  $\jump \colon \nat \to \nat$ that gives us how long we stay inside
  some component as $\jump \colon i \mapsto j-i$, where $j$ is the
  largest number such that $C(q_j)=C(q_i)$.

  It is easy to see by the definitions of the transitions in
  $\delta_\rightarrow$ and $\delta_\circlearrowright$, that the run
  \[ \big(p_0,\min(N^2,\jump(0)),0\big) \;\;\rightarrow\;\; \cdots \;\;\rightarrow\;\;
    \big(p_n,\min(N^2,\jump(n)),0\big) \] is an accepting run of $A$, where
  $p_i$ is $q_i$ if $c_i<N^2$ and $[q_i]$ otherwise.  We note that the
  counter $d$ is always zero, as we do not use any transitions from
  $\delta_\curvearrowright$. The transitions in
  $\delta_\curvearrowright$ are only there to satisfy the consistent
  jumps property.  This shows $L(A_L) \subseteq L(A)$.

  Towards the lemma statement, it remains to show that
  $L(A) \subseteq L(A_L)$. Let therefore $w=a_1 \cdots a_n$ be some
  string in $L(A)$,
  $(p_0,c_0,d_0) \rightarrow \cdots \rightarrow (p_n,c_n,d_n)$ be an
  accepting run of $A$, and $q_0\rightarrow \cdots \rightarrow q_n$ be the unique run of
  $A_L$ on $w$.

  We now show by induction on $i$ that there are states
  $\hat{q}_1,\dots,\hat{q}_n$ in $Q_L$ such that the following claim
  is satisfied. The claim easily yields that $q_n \in F_L$, as both
  counters have to be zero for the word to be accepted.
  \[  L_{\hat{q}_i} \cap a_{i+1} \cdots a_{i+d_i}\Sigma^* \subseteq L_{q_i}\; \text{ and } \;\hat{q}_i \in
    \begin{cases}
      \{p_i\} & \text{if } c_i=d_i=0 \\
      [p_i] & \text{if } c_i+d_i>0 \text{ and } p_i \in Q_L \\
      p_i & \text{if } c_i+d_i>0 \text{ and } p_i \in [Q_L] \\
    \end{cases}
  \]

  The base case $i=0$ is trivial by the definition of $I$. We now
  assume that the induction hypothesis holds for $i$ and are going to
  show that it holds for $i+1$. Let $\rho=(p_{i-1},a_i,P;p_i,U)$ be
  the transition used to read $a_i$.  We distinguish several cases
  depending on $\rho$.

  Case $\rho \in \delta_\rightarrow$: In this case, $c_i=0$ by the
  definition of $\delta_\rightarrow$. Therefore, the claim for $i+1$
  follows with $\hat{q}_{i+1}=p_{i+1}$, as $\hat{q}_i=p_i$ by the
  induction hypothesis and $(p_i,a,p_{i+1}) \in \delta_L$ by the
  definition of $\delta_\rightarrow$.

  \medskip

  Case $\rho \in \delta_\circlearrowright^2$: We note that
  $p_i,p_{i+1} \in [Q_L]$. The claim for $i+1$ follows with
  $\hat{q}_{i+1}=\delta(\hat{q}_i,a_{i+1})$ using
  $C(q')=C(\delta(q',a_{i+1})$ for all $q' \in [p_i]$ (by
  Observation~\ref{obs:component}), $C(p_i)=C(p_{i+1})$ (by definition of
  $\delta_\circlearrowright$), and $\hat{q}_i \in p_i$ (by the
  induction hypothesis).

  \medskip

  Case $\rho \in \delta_\circlearrowright^3$: We want to show that
  $L_{p_{i+N^2}} \subseteq L_{q_{N^2}}$ establishing the claim
  directly for the position $i+N^2$ using $\hat{q}_{i+N^2}=p_{i+N^2}$.
  Therefore, we first want to apply Lemma~\ref{lem:sameState} to show
  that $\delta(\hat{q}_i, a_{i+1} \cdots a_{i+N^2})=p_{i+N^2}$. The
  preconditions of the lemma require us to show that (i)
  $C(\hat{q}_i)=C(p_i)$, (ii) $C(p_i)=C(p_{i+N^2})$, and (iii)
  $C(\hat{q}_i)=C(\delta_L(\hat{q}_i,a_{i+1} \cdots a_{i+N^2}))$.
  Precondition (i) is given by the induction hypothesis,
  precondition~(ii) is by the definition of
  $\delta_\circlearrowright$, i.e., that all transitions in
  $\delta_\circlearrowright$ are inside the same component of $A_L$,
  and precondition~(iii) is by the fact that each transition in
  $\delta_\circlearrowright$ has a corresponding transition in
  $\delta_L$ that stays in the same component. Therefore, we can
  actually apply Lemma~\ref{lem:sameState} to conclude that
  $\delta(\hat{q}_i, a_{i+1} \cdots a_{i+N^2})=p_{i+N^2}$.  As we
  furthermore have that $L_{\hat{q}_i} \cap a_{i+1} \cdots a_{i+d_i}\Sigma^* \subseteq L_{q_i}$  by the
  induction hypotheses, we can conclude that
  $L_{p_{i+N^2}} \subseteq L_{q_{N^2}}$. This establishes the claim for
  position $i+N^2$ using $\hat{q}_{i+N^2}=p_{i+N^2}$. As we only need
  the claim for position $n$ (and not for all smaller positions), we
  can continue the induction at position $i+N^2$. Especially there is
  no need to look at the case where
  $\rho \in \delta_\circlearrowright^1$.

  \medskip

  Case $\rho \in \delta_\curvearrowright$: By the definition of
  $\delta_\curvearrowright$, we have that $p_i,p_{i+1} \in [Q_L]$.
  Furthermore, there are transitions $(p_i,a_{i+1},p')$ and
  $(p'',a_{i+1},p_{i+1})$ in $\delta_L$ such that $C(p')=C(p_i)$,
  $C(p'')=C(p_{i+1})$, and $p' \leadsto p''$.  This (and the fact that
  $\hat{q}_i \in p_i$ by the induction hypothesis) allows us to apply
  Observation~\ref{obs:component}, which yields
  $\delta(\hat{q}_i,a_{i+1}) \in C(p_i)$.  From $p' \leadsto p''$ and
  $\hat{q}_i \in C(p')$ we can conclude that $\hat{q}_i \leadsto p''$.
  We now can apply Lemma~\ref{lem:Bagan9} that gives us
  $L_{p''} \cap L_{p''}^{a_{i+1}}\Sigma^* \subseteq L_{\hat{q}_i}$.

  Now we argue that the subword $a_{i+2} \cdots a_{i+N^2+1}$ is in
  $L_{p''}^{a_{i+1}}$.  By the definition of
  $\delta_\curvearrowright$, we have $d_{i+1}=N^2$, enforcing that the
  next $N^2$ transitions are all from $\delta_\circlearrowright^2$, as
  these are the only transitions that allow $d>0$ in the
  precondition. Applying Observation~\ref{obs:component} $N^2$ times
  yields that
  $\delta(p'',a_{i+2} \cdots a_{i+N^2+1}) \in
  C(p'')$ and therefore
  $a_{i+2} \cdots a_{i+N^2+1} \in L_{p''}^{a_{i+1}}$. Using this and
  $L_{p''} \cap L_{p''}^{a_{i+1}}\Sigma^* \subseteq L_{\hat{q}_i}$,
  we get that
  $L_{\delta(p'',a_{i+1})} \cap a_{i+2}\cdots a_{i+N^2+1}\Sigma^* \subseteq
  L_{\delta(\hat{q}_i,a_{i+1})}$ yielding the claim for $i+1$.
  This concludes the proof of the lemma.
\end{proof}

We now continue with the rest of the proof of Theorem~\ref{thm:ttractNFA}.
\begin{proof}[Proof of Theorem~\ref{thm:ttractNFA}]
  We show (1) $\Rightarrow$ (5) $\Rightarrow$ (3) $\Rightarrow$ (2) $\Rightarrow$ (1) and (5) $\Rightarrow$ (4) $\Rightarrow (2)$.

  (1) $\Rightarrow$ (5): Holds by Lemma~\ref{lem:memoryless}.

  \medskip

  (5) $\Rightarrow$ (3) and (4) $\Rightarrow$ (2): Let
  $A=(Q,I,c,\delta,F,\ell)$ be a detainment automaton with consistent
  jumps.
  We compute an equivalent NFA
  $A'=(Q \times \{0,\dots,\ell\},\delta',I \times \{0\},F \times \{0\})$
  in the obvious way, i.e., $((p,i),a,(q,j)) \in \delta'$ if and only
  if $A$ can go from configuration $(p,i)$ to configuration $(q,j)$
  reading symbol $a$.
  By the definition of detainment automata, we get that the \components of $A'$ are
  \[
    \{\;C \times \{0\} \;\mid\; C \text{ is a \component of } A\;\}
  \]
  This directly shows that $A'$ only has memoryless components if $A$ only has memoryless components.

  To prove the left-synchronizing containment property, we choose $n=\ell$.
  Let now $(q_1,c_1),(q_2,c_2) \in Q \times \{0,\dots,\ell\}$,
  $a \in \Sigma$, and $w_1',w_2' \in \Sigma^*$ be such that
  $(q_1,c_1) \leadsto (q_2,c_2)$,
  $w_1=aw_1' \in \loopwords((q_1,c_1))$, and
  $w_2=aw_2' \in \loopwords((q_2,c_2))$. We have to show that
\begin{equation}
  w_2^{n}L_{(q_2,c_2)} \;\;\subseteq\;\; L_{(q_1,c_1)}\;.\label{eq:CNFAtoNFA}
\end{equation}
We distinguish two cases. If $q_1$ and $q_2$ are in the same
component, we know that there is a transition
$(q_1,a,\true;q_3;c:=c-1) \in \delta$, as $A$ has consistent jumps.
Therefore, there is a transition $((q_1,0),a,(q_2,0)) \in \delta'$,
which directly yields~\eqref{eq:CNFAtoNFA}.

If $q_1$ and $q_2$ are in different components, then there is a
transition $(q_1,a,c=0;q_3;c:=k) \in \delta$, as $A$ has consistent
jumps. Therefore, there is a transition
$((q_1,0),a,(q_2,k)) \in \delta'$ for some $k \in [0,\ell]$.  We have
$w_2 \in \loopwords(q_2)$. The definition of detainment automata
requires that every transition inside a component---thus every
transition used to read $w_2$ using the loop---is of the form
$(p,a,\true;q,c:=c-1)$, i.e., it does not have a precondition and it
decreases the counter by one. Therefore in $A'$, we have that
$\delta'((q_2,k),w^\ell) \supseteq \delta((q_2,0),w^\ell)$.
This concludes the proof of (5) $\Rightarrow$ (3) and (4) $\Rightarrow$ (2)

\medskip

(5) $\Rightarrow$ (4) and (3) $\Rightarrow$ (2): Trivial.

\medskip

(2) $\Rightarrow$ (1): Let $A=(Q,\Sigma,\delta,I,F)$ be an NFA
satisfying the \lscp and $A_L$ be the minimal DFA equivalent to
$A$. We show that $A_L$ satisfies the \lscp establishing~(1).

Let $M$ be the left synchronizing-length of $A$ and $q_1,q_2 \in Q_L$ be
states of $A_L$ such that
\begin{itemize}
\item $q_1 \leadsto q_2$; and
\item there are words $w_1\in \loopwords(q_1)$ and $w_2\in \loopwords(q_2)$ that start with the same
  symbol $a$.
\end{itemize}
We need to show that there exists an $n \in \nat$ with
$w_2^{n}\cL_{q_2} \subseteq \cL_{q_1}$.  Let $w$ be a word such that
$\delta(q_1,w)=q_2$. Let $P_1 \subseteq Q$ be a state of the powerset
automaton of $A$ with $\cL_{P_1}=\cL_{q_1}$ and let
$P_2=\delta(P_1,ww_2^*)$ be the state in the powerset automaton of $A$ that
consists of all states reachable from $P_1$ reading some word from $ww_2^*$.

It holds that $\cL_{P_2}=\cL_{q_2}$, as
$\delta(q_1,ww_2^*)=q_2$ and $\cL_{q_1}=\cL_{P_1}$.

We define
\begin{align*}
P_2'\quad&=\quad\{\;p \in P_2 \mid w_2^i \in \loopwords(p) \text{ for some } i>0\;\} \\
P_2'' \quad&=\quad\delta(P_1,w_2^{|A|})
\end{align*}

We obviously have $P_2'' \subseteq P_2' \subseteq P_2$. Furthermore, we have
\[ \cL_{P_2} \;\;=\;\; \cL_{q_2} \;\;=\;\;
  \cL_{\delta(q_2,w_2^{|A|})} \;\;=\;\; \cL_{P_2''} \] The second
equation is by $\delta(q_2,w_2^{|A|})=q_2$. We can conclude that
$\cL_{q_2}=\cL_{P_2'}$.

Let $\rho\colon Q \to Q$ be a function that selects for every state
$p_2 \in P_2'$ a state $p_1 \in P_1$ such that $p_1 \leadsto p_2$. By
definition of $P_2'$, such states exist.  Using the fact that $A$
satisfies the \lscp, we get that
$w_2^M\cL_{p_2} \subseteq \cL_{\rho(p_2)}$ for each $p_2 \in P_2$. We
can conclude
\[
    w_2^M\cL_{q_2} \quad =\quad w_2^M\cL_{P_2'} \quad=\quad\bigcup\limits_{p_2 \in P_2'} w_2^M\cL_{p_2} \quad\subseteq\quad
    \bigcup\limits_{p_2 \in P_2'} \cL_{\rho(p_2)} \quad\subseteq\quad\cL_{P_1}\quad =\quad \cL_{q_1}
  \] and therefore
  $w_2^{|A|+M}\cL_{q_2} \subseteq \cL_{q_1}$.
  So $A_L$ satisfies the \lscp with $n = M$, where $M$ is the left synchronizing-length of $A$.
  This concludes the proof for (2) $\Rightarrow$ (1) and thus the proof of the theorem.
\end{proof}

\subsection{Regular Simple Path Queries}\label{subsec:sptrac}
Bagan et al.~\cite{BaganBG-jcss20} analyzed the problem of computing all simple paths that satisfy a given regular path query.
They
introduced %
the class \ctract, which characterizes the class
of regular languages $L$ for which the \emph{regular simple path query (RSPQ)} problem
is tractable. %
\begin{thmC}[{\cite[Theorem~3]{BaganBG-jcss20}}]
  Let $L$ be a regular language.
  \begin{enumerate}[(1)]
  \item If $L$ is finite, then $\rspq(L) \in \text{AC}^0$.
  \item If $L \in \ctract$ and $L$ is infinite, then $\rspq(L)$ is NL-complete.
  \item If $L \notin \ctract$, then $\rspq(L)$ is NP-complete.
  \end{enumerate}
\end{thmC}

\noindent
One characterization of \ctract is the following (Theorem 6 in~\cite{BaganBG-jcss20}):
\begin{thm}\label{thm:sptract}
  \ctract is the set of regular languages $L$
  such that there exists an $i \in \nat$ for which the following holds:   for all $w_\ell, w, w_r \in \Sigma^*$ and $w_1,w_2 \in
  \Sigma^+$ we have that, if $w_\ell w_1^i w w_2^i w_r \in L$, then $w_\ell
  w_1^i w_2^i w_r \in L$.
\end{thm}

Comparing the characterization with Definition~\ref{def:leftsyncpower}, we see that Definition~\ref{def:leftsyncpower}
imposes an extra ``synchronizing'' condition on $w_1$ and $w_2$, namely that they share the same first (or last) symbol.
We therefore have the following observation:
\begin{obs}\label{obs:sptract-in-ttract}
  The class \ctract is contained in \etract.
\end{obs}

\subsection{An algebraic Characterization of \texorpdfstring{\ttract}{T-tract} and \texorpdfstring{\ctract}{SP-tract}}\label{subsec:comparison}
We now provide an algerbaic characterization of \ttract and
\ctract. We use this characterization for two things: First, we use
the characterizations to fully classify the expressiveness of both
classes with respect to some well known fragments of first order
logic.  The results are depicted in
Figure~\ref{fig:allClasses}. Later, in Section~\ref{sec:closureproperties}, we will
conclude a bunch of closure properties for both classes. These
properties follow from Observation~\ref{obs:ne-varieties}, which is
the only result from this subsection that is used outside of it.

\begin{figure}
    \centering
    \begin{tikzpicture}
    \draw (3.5,4.8) node[anchor=west] {aperiodic languages ($=\FO[<]$)};
    \draw (9,4.8) node[anchor=west] {$(ac^*bc^*)^*$};
    \draw[rounded corners] (1,4.3) rectangle (13, 8) {};

    \draw (3.5*1.9,6.7) ellipse (0.4cm and .3cm);%
    \draw (3.5*1.9,6.7) node {$\text{\dc}$};

    \draw (3*2,6.7) ellipse (1.2cm and .5cm);%
    \draw (3*1.9,6.9) node {$\text{\sptract}$};
    \draw (3*1.9,6.5) node {$a$};
    \node at (3*2,6) {$a^*bc^*$};

    \draw (7.5,6.5) ellipse (3cm and 1cm);%
    \draw (8.8,6.7) node {$\text{\ttract}$};
    \node at (8.8,6) {$(ab)^*$};

    \draw (4.5,6.5) ellipse (3cm and 1cm);%
    \draw (3.2,6.7) node {$\textbf{FO}^2[<]$};
    \node at (3.3,6) {$a^*ba^*$};

      \draw[rounded corners] (1.3,5.3) rectangle (12.7, 7.7) {};
      \node at (11.5,6.7) {$\textbf{FO}^2[<,+1]$};
      \node at (11.5,6) {$a^*ba^*(cd)^*$};

    \end{tikzpicture}
    \caption{Expressiveness of subclasses of the aperiodic languages
    }%
    \label{fig:allClasses}
\end{figure}

We refer the reader to the book~\cite{Pin-handbook97} for a general overview of syntactic semigroups and the different hierarchies. We use the following notation.
The syntactic preorder of a language $L$ of $\Sigma^*$ is the relation $\leq_L$ defined on $\Sigma^*$ by $x \leq_L y$ if and only if for all $u,v \in \Sigma^*$ we have $u x v \in L \Rightarrow u y v \in L$.
The syntactic congruence of $L$ is the associated equivalence relation $\sim_L$ defined by $x \sim_L y$ if and only if $x\leq_L y$ and $y\leq_L x$. The quotient $\Sigma^+\slash{}\sim_L$ ($\Sigma^*\slash{}\sim_L$) is called the syntactic semigroup (monoid) of $L$.
A word $e \in \Sigma^*$ is \emph{idempotent} if $e^2 = e$. Given a finite semigroup $S$, it is folklore that there is an integer $\omega(S)$ (denoted by $\omega$ when $S$ is understood) such that for all $s \in S$, $s^\omega$ is idempotent. More precisely, $s^\omega$ is the limit of the Cauchy sequence $(s^{n!})_{n\geq 0}$.

Let \textbf{SP} denote the variety of semigroups defined by the profinite inequalities $x^\omega u y^\omega \leq
x^\omega y^\omega$. Let \textbf{T} denote the variety of semigroups defined by the set of profinite inequalities
$(xy)^\omega u (xz)^\omega \leq (xy)^\omega (xz)^\omega$. Note that, by choosing $x = y$ and/or $x = z$, this last inequality implies that the profinite inequalities $(x)^\omega u (xz)^\omega \leq (x)^\omega(xz)^\omega,
(xy)^\omega u (x)^\omega \leq (xy)^\omega(x)^\omega$ and $(x)^\omega u (x)^\omega \leq (x)^\omega$ must also be satisfied.

\begin{thm}\label{theo:algebraic-characterization-tract}\hfill
	\begin{enumerate}[(1)]
		\item A regular language $L \in \Sigma^+$ is in
		\ctract if and only if its syntactic semigroup belongs to~\textbf{SP}.
		\item A regular language $L \in \Sigma^+$ is in \ttract if and only if its syntactic semigroup belongs to~\textbf{T}.
	\end{enumerate}
\end{thm}
\begin{proof}
	Item (1) follows from Theorem~\ref{thm:sptract} and the observation that if there exists an $i$ for which Theorem~\ref{thm:sptract} holds, then it also holds for each $i' \geq i$. This can easily be seen by choosing $w'_\ell =w_\ell w_1^{i'-i}$ and $w'_r = w_2^{i'-i}w_r$.
	Item (2) follows from Theorem~\ref{theo:equivalence}, Definition~\ref{def:leftsyncpower} and the paragraph after the definition.
\end{proof}

\begin{obs}\label{obs:ne-varieties}
The theorem immediately implies that \sptract and \ttract are varieties of semigroups and ne-varieties~\cite{Pin-handbook97,PinS-ita05}.
\end{obs}

We now fully classify the expressiveness of \ttract and \sptract compared to
yardstick classes such as \dc, $\FO^2[<]$, and
$\FO^2[<,+1]$ (see also
Figure~\ref{fig:allClasses}).
Here, $\FO^2[<]$ and $\FO^2[<,+1]$ are the two-variable restrictions of
$\FO[<]$ and $\FO[<,+1]$ over words, respectively. By $\FO[<,+1]$ we mean the
first-order logic with unary predicates $P_a$ for all $a \in \Sigma$ (denoting
positions carrying the letter $a$) and the binary predicates $+1$ and $<$
(denoting the successor relation and the order relation among positions).
The logic $\FO[<]$ is $\FO[<,+1]$ without the successor predicate.

We use the characterizations from
Theorem~\ref{theo:algebraic-characterization-tract} to classify $\sptract$ and
$\ttract$ \mbox{wrt.} the Straubing-Thérien
hierarchy~\cite{Straubing-tcs81,Therien-tcs81}) and the dot-depth hierarchy
(also known as Brzozowski hierarchy~\cite{CohenB-jcss71}). Both hierarchies are
particular instances of concatenation hierarchies, which means that they can be
built through a uniform construction scheme. Pin~\cite{Pin-17} summarized
numerous results and conjectures around these hierarchies.

Thomas~\cite{Thomas-jcss82} showed that the dot-depth hierarchy corresponds, level by level, to the
quantifier alternation hierarchy of first-order formulas, defined as follows.
A formula is a $\Sigma_n$-formula if it is equivalent to a formula $Q(x_1, \ldots, x_k) \varphi$, where $\varphi$ is quantifier free and $Q(x_1, \ldots, x_k)$ is a sequence of $n$ blocks of quenatifiers such that the first block contains only existental quantifiers.
The class $\Sigma_n$ is the class of languages which can be defined by $\Sigma_n$-formulas. The class $\Pi_n$ is defined by starting with a block of universal instead of existential quantifiers. A language $L$ is in $\Pi_1[<, +1]$ if and only if its complement $L^c$ is in $\Sigma_1[<, +1]$.
The class of downward closed languages $\dc$ is exactly the class $\Pi_1[<]$.

\begin{thm}\leavevmode\label{theo:between-pi1s}
	$\dc \subsetneq \sptract \subsetneq \ttract \subsetneq \Pi_1[<,+1]$
\end{thm}
\begin{proof}
	We first show $\dc \subsetneq \sptract$.
	As \dc is definable by simple regular expressions, we have for each downward closed language $L$ that $w_\ell w_1^i w w_2^i w_r \in L$ implies $w_\ell w_1^i w_2^i w_r \in L$ for every integer $i\in \nat$ and all words $w_\ell, w_1, w, w_2, w_r \in \Sigma^*$. Therefore, $L \in \ctract$ by Theorem~\ref{thm:sptract}.
	The language $\{a\}$ is not downward closed, but in \ctract using Theorem~\ref{thm:sptract} with $i=1$.

	The subset relation $\sptract \subseteq \ttract$ was already
        obserbed earlier (Observation~\ref{obs:sptract-in-ttract}) and
        $a^*bc^*$ is a language in \ttract which is not in \sptract,
        showing that the containment is strict.

	We now prove that $\ttract$ is strictly contained in $\Pi_1[<,+1]$. The class $\Sigma_1[<,+1]$ is defined by the equation
        $x^\omega u x^\omega \geq x^\omega$, see Pin and
        Weil~\cite{PinW-Comm02}. As $\Pi_1[<,+1]$ is the dual variety
        of $\Sigma_1[<,+1]$, it is defined by
        $x^\omega u x^\omega \leq x^\omega$.  By
        Theorem~\ref{theo:algebraic-characterization-tract} it
        immediately follows that \ttract is in the class satisfied by
        $x^\omega u x^\omega \leq x^\omega$ and therefore
        $\ttract \subseteq \Pi_1[<,+1]$. On the other hand,
        $(a+b)^* a (a+b)^*$ is an example of a language definable in
        $\Pi_1[<,+1]$ which is not in \ttract.
\end{proof}

So \sptract and \ttract are between $\Pi_1[<]$ and
$\Pi_1[<,+1]$.

While \sptract and \ttract behave similar when the number of alternations of a first-order formula is restricted, restricting the number of variables ($\FO^2$) leads to a different behavior:

\begin{thm}\leavevmode\label{theo:classification}
  \begin{enumerate}[(a)]
  \item $\sptract \subsetneq \FO^2[<]$
  \item \ttract and $\FO^2[<]$ are incomparable
  \item $\ttract \subsetneq \FO^2[<,+1]$
  \end{enumerate}
\end{thm}
\begin{proof}
  We first show~(a).
  Th\'erien and Wilke~\cite{TherienW-stoc98} proved that \textbf{DA}=$\textbf{FO}^2[<]$, where \textbf{DA} is defined by the identity
   $(xyz)^\omega y (xyz)^\omega = (xyz)^\omega$.
   Thus we only have to prove that each syntactic semigroup of a language in \sptract satisfies this identity. Let $L \in \sptract$.
   By Theorem~\ref{theo:algebraic-characterization-tract}, it immediately follows that the syntactic semigroup of $L$ satisfies $(xyz)^\omega y (xyz)^\omega \leq (xyz)^\omega$. Thus it remains to show that there exists an $n'$ such that for each $n \geq n'$ and all $u, v,x,y,z \in \Sigma^*$ it holds that:
     \[u (xyz)^n y (xyz)^n v \in L  \;\;\text{ if }\;\; u (xyz)^n v \in L\;. \]

    For this direction, we use that Bagan et \mbox{al.}~\cite[Theorem 6]{BaganBG-jcss20} give a definition of \sptract in terms of regular expressions, showing that each \component can be represented as $(A^{\geq k} + \varepsilon)$ for some set $A \subseteq \Sigma$ and $k\in \nat$.
    So if there is $xyz \in \Sigma^*$ with $ u (xyz)^M v \in L$ for some $u,v \in \Sigma^*$, then we also have
    $u (xyz)^M (\alphabet(xyz))^* (xyz)^M v \subseteq L$, where $\alphabet(x)$
    denotes the set of symbols $x$ uses. Thus we especially have $u (xyz)^M y
    (xyz)^M v \in L$, which proves the other direction. The same holds for each $M' \geq M$. This concludes the proof
    of~(a).

    Statement~(b)
    simply follows from the facts that the language $a^*ba^*$ is in
    $\FO^2[<]$ but not in \ttract whereas the language $(ab)^*$ is in
    \ttract but not in $\FO^2[<]$.

   It remains to prove~(c), which follows from Theorem~\ref{theo:between-pi1s} as $\Pi_1[<,+1]$ is a subset of the 1st level of the dot-depth hierarchy, which in turn is a subset of $\FO^2[<,+1]$. The language $a^*ba^*$ is an example of a language in $\FO^2[<,+1]$ and not in \ttract.
\end{proof}

\begin{prop}\leavevmode\label{prop:hierarchy}
	$\sptract$ is in $\mathcal{V}_{3/2}$, the 3/2th level of the Straubing-Thérien hierarchy.
\end{prop}
\begin{proof}
	Let $L \in \ctract$. The 3/2th level of the Straubing-Thérien hierarchy is defined by the profinite inequality $x^\omega \leq x^\omega y x^\omega$ where $\alphabet(x) = \alphabet(y)$~\cite[Theorem 8.9]{Pin-handbook97}. This means that we have to show that there exists an $n'$ such that for all $n \geq n'$ and words $w_\ell, w_r$ it holds: if $w_\ell x^n w_r$ in $L$, then also $w_\ell x^n y x^n w_r$ in $L$. We can easily see that every language in \sptract satisfies this: The \components have the form $(A^{\geq k} + \varepsilon)$ for some set of symbols $A$ by the definition of \ctract in terms of regular expressions, see~\cite[Theorem 6]{BaganBG-jcss20}. Therefore, the implication immediately holds for all $y$ with $\alphabet(y) = \alphabet(x)$.
\end{proof}

\section{The Trichotomy}\label{sec:trichotomy}
This section is devoted to the proof of the following theorem.
\begin{thm}\label{theo:trichotomy}
  Let $L$ be a regular language.
  \begin{enumerate}[(1)]
  \item If $L$ is finite then $\rtq(L) \in \text{AC}^0$.
  \item If $L \in \etract$ and $L$ is infinite, then $\rtq(L)$ is NL-complete.\label{theo:trichotomy-ttract}
  \item If $L \notin \etract$, then $\rtq(L)$ is NP-complete.\label{theo:trichotomy-np}
  \end{enumerate}
\end{thm}

\subsection{Finite Languages}\label{sec:finitelanguages}
We now turn to proving Theorem~\ref{theo:trichotomy}. We start with Theorem~\ref{theo:trichotomy}(1). Clearly, we can express every finite language $L$ as a 
FO-formula. Since we can also test in FO that no edge $e$ is used more than once,
the multigraphs for which $\rtq(L)$ holds are FO-definable. By Immerman~\cite{Immerman-siamcomp88}, this implies that
$\rtq(L)$ is in AC$^0$.

\subsection{Languages in \texorpdfstring{\ttract}{Ttract}}
We now sketch the proof of Theorem~\ref{theo:trichotomy}\ref{theo:trichotomy-ttract}.
We note that we define several
concepts (trail summary, local edge domains, admissible trails) that have a
natural counterpart for simple paths in Bagan et al.'s proof of the trichotomy 
for simple paths~\cite{BaganBG-jcss20}.
 However, the
underlying proofs of the technical lemmas are quite different.
 For
instance, \components of languages in \ctract behave similarly
to $A^*$ for some $A \subseteq \Sigma$, while \components of languages in \ttract
are significantly more complex. Furthermore, the trichotomy for trails leads to
a strictly larger class of tractable languages.

For the remainder of this section, we fix the constant $K=N^2$.

We will show that in the case where $L$ belongs to \ttract, we can identify a
number of edges that suffice to check if the path is (or can be transformed
into) a trail that matches $L$. This number of edges only depends on $L$ and is
therefore constant for the $\rtq(L)$ problem. These edges will be stored in a
\emph{summary}. We will define summaries formally and explain how to
use them to check whether a trail between the input nodes that matches $L$
exists.
To this end, we need a few definitions.

\begin{defi}\label{def:longrun}
	Let $p = e_1 \cdots e_m$ be a path and $r =q_0 \rightarrow \cdots \rightarrow q_m$ the run
	of $A_L$ over $\lab(p)$. For a set $C$ of states of $A_L$, we denote
	by $\Left_C$ the first edge $e_i$ with
	$q_{i-1} \in C$ and by $\Right_C$ the last edge $e_j$ with $q_j \in C$.
	A component $C$ of $A_L$ is a \emph{long run component of $p$} if
	$\Left_C$ and $\Right_C$ are defined and
	$|p[\Left_C,\Right_C]| > K$.
\end{defi}

Next, we want to reduce the amount of information that we require for trails.
The synchronization property, see Lemma~\ref{lem:sameState}, motivates the use of \emph{summaries}, which we define next.

\begin{defi}
  Let $\Cuts$ denote the set of components of $A_L$ and $\Abbrv =\Cuts \times (V
  \times Q) \times E^K$. A \emph{component abbreviation} $(C,(v,q),e_K \cdots
  e_1) \in \Abbrv$ consists of a component $C$, a node $v$ of $G$ and state $q
  \in C$ to start from, and $K$ edges $e_K \cdots e_1$. A trail $\pi$
  \emph{matches} a component abbreviation, denoted $\pi \models (C,(v,q),e_K
  \cdots e_1)$, if $\delta_L(q,\pi) \in C$, it starts at $v$, and its suffix is
  $e_K \cdots e_1$. Given an arbitrary set of edges $E'$, we write $\pi \models_{E'} (C,(v,q),e_K \cdots e_1)$ if $\pi
  \models (C,(v,q),e_K \cdots e_1)$ and all edges of $\pi$ are from $E' \cup
  \{e_1,\dots,e_K\}$.
  For convenience, we write $e \models_\emptyset e$.

  If $p$ is a trail, then the \emph{summary $S_p$} of $p$ is the
  sequence obtained from $p$ by replacing, for each long run component
  $C$, the subsequence $p[\Left_C,\Right_C]$ by the abbreviation
  $(C,(v,q),p_\mathsf{suff})$, where $v$ is the source node of the edge $\Left_C$,  $q$ is the state in which $A_L$ is
   immediately before reading $\Left_C$, and $p_\mathsf{suff}$ is the suffix of
  length $K$ of $p[\Left_C,\Right_C]$.
\end{defi}
We note that the length of a summary is always bounded by $O(N^3)$,
i.e., a constant that
depends on $L$. Indeed, $A_L$ has at most $N$ components and, for each of them,
we store at most $K+3$ many things (namely, $C, v, q$, and $K$ edges).
Our goal is to find a summary $S$ and replace all abbreviations
with matching pairwise edge-disjoint trails which do not use any other edge in $S$,
because this results in a trail that matches $L$. However, not every sequence of
edges and abbreviations is a summary, because a summary needs to be obtained from a trail. So, we will work with candidate summaries instead.
\begin{defi}\label{def:candidatesummary}
  A \emph{candidate summary $S$} is a sequence of the form
  $S= \alpha_1 \cdots \alpha_m$ with $m\leq N$
  where each $\alpha_i$
  is either (1) an edge $e \in E$ or (2) an
  abbreviation $(C,(v,q),e_K \cdots e_1) \in \Abbrv$.
  Furthermore, all components in $S$ are distinct and each edge $e$ occurs at most once.
  A path $p$ that is derived from $S$ by replacing each
  $\alpha_i \in \Abbrv$ by a trail $p_i$ such that
  $p_i \models \alpha_i$ is called a \emph{completion} of the
  candidate summary $S$.
\end{defi}
The following corollary is immediate from the definitions and
Lemma~\ref{lem:sameState}, as the lemma ensures that the state after
reading $p$ inside a component does not depend on the whole path but only on the labels
of the last $K$ edges, which are fixed.
\begin{cor}\label{cor:Bagan10}
  Let $L$ be a language in \ttract. Let $S$ be the summary of a trail $p$ that
  matches $L$ and let $p'$ be a completion of $S$. Then, $p'$ is a path that
  matches $L$.
\end{cor}

Together with the following lemmas, Corollary~\ref{cor:Bagan10} can be used to
obtain an nondeterministic logarithmic space algorithm\footnote{That is, a nondeterministic Turing Machine with read-only input and write-only output that only uses $O(\log{n})$ space on its working tapes.} that gives us a
completion of a summary $S$.
The lemma heavily relies
on other results on the structure of components in $A_L$.
\begin{lem}\label{lem:return-shortest-path-nl}
	There exists a nondeterministic logarithmic space algorithm that, given a
	directed graph $G$ and nodes $s$ and $t$, outputs a shortest path from $s$ to
	$t$ in $G$.
\end{lem}
\begin{proof}
	We show that Algorithm~\ref{alg:immermann-szelepscenyi} can output a shortest
	path in nondeterministic logarithmic space. Recall that nondeterministic algorithms with output either give up, or produce a correct output and that at least one computation does not give up. We note that Algorithm~\ref{alg:immermann-szelepscenyi} is a mixture of the Immermann-Szelepscényi Theorem~\cite{Immerman-siamcomp88,Szelepcsenyi-acta88} and reachability.
	To this end, $S(k)$ denotes the set of nodes reachable from $s$ with $k$ edges. Using the algorithm given by Immermann~\cite{Immerman-siamcomp88} and Szelepscényi~\cite{Szelepcsenyi-acta88} to show that non-reachability is in \nlogspace, we can find in lines~\ref{alg:immermann:1}--\ref{alg:immermann:before-reachability} the smallest $n$ such that a path from $s$ to $t$ of length $n$ but none of length $n-1$ exists. Indeed, we only added a test in line~\ref{alg:immermann:test-t} to find the smallest $k$ for which $t \in S(k)$---this $k$ is the length of a shortest path from $s$ to $t$.
	After line~\ref{alg:immermann:continue-after-n-found} we then use the smallest $k$ (which we name $n$) together with a standard reachability algorithm to nondeterministically output a path of this length. (If we are only interested in the length of a shortest path, we can return $n$ instead.) We note that one can easily change the algorithm to avoid outputting edges of paths that will give up. This would require an extra test if there exists a path of length $n-p$ from $w_p$ to $t$ before outputting the edge from $w_{p-1}$ to $w_p$. We omitted this extra test for readability (and because at this point we know that there is a solution and non-deterministic algorithms will always return the correct output).

	That Algorithm~\ref{alg:immermann-szelepscenyi} runs in nondeterministic
	logarithmic space follows from the Immermann-Szelepscényi Theorem and reachability being in \nlogspace.
\end{proof}
\begin{algorithm}[p]{}
	\DontPrintSemicolon%
	\KwIn{A directed graph $G=(V,E,\edge)$, nodes $s,t$ in $G$, $s \neq t$}
	\KwOut{A shortest path from $s$ to $t$ in $G$ or ``no'' if no path from $s$ to $t$ exists}

	$n\gets -1$\label{alg:immermann:1}\Comment*[r]{$n$ will be the length of a shortest path from $s$ to $t$}
	$|S(0)|\gets 1$ \;
	\For(\Comment*[f]{Compute $|S(k)|$ from $|S(k-1)|$}){$k = 1, 2, \ldots, |V|-1$}{
		$\ell \gets 0$\;
		\ForEach(\Comment*[f]{Test if $u \in S(k)$}){$u \in V$}{
			$m\gets 0$\;
			\textit{reply} $\gets \false$\;
			\ForEach(\Comment*[f]{Test if $v \in S(k-1)$}){$v \in V$}{
				$w_0\gets s$\;
				\For{$p=1,\ldots, k-1$}{
					guess a node $w_p$ \;
					\If{$(w_{p-1},w_p)$ is not an edge in $G$\label{alg:immermann:testedge1}}{\textbf{give up}}}
				\If{$w_{k-1}\neq v$}{\textbf{give up}}
				$m\gets m+1$\;
				\If{$(v,u)$ is an edge in $G$\label{alg:immermann:testedge2}}{
					\textit{reply} $\gets \true$\;
					\If{$u = t$\label{alg:immermann:test-t}}{$n \gets k$\;
						continue in line~\ref{alg:immermann:continue-after-n-found}
					}
				}
			}
			\If{$m < |S(k-1)|$}{\textbf{give up}}
			\If{reply $= \true$}{
				$\ell \gets \ell+1$
			}
		}
		$|S(k)|\gets \ell$
	}
	\Return ``no''\label{alg:immermann:before-reachability} \Comment*[r]{$t\notin S(k)$ for any $k$} 
	$w_0\gets s$\label{alg:immermann:continue-after-n-found}\;
	\For{$p=1,\ldots, n$}{
		guess a node $w_p$\label{alg:immermann:guess-node} \;
		\If{$(w_{p-1},w_p)$ is not an edge in $G$\label{alg:immermann:testedge3}}{\textbf{give up}}
		output an edge from $w_{p-1}$ to $w_p$ in $G$\label{alg:immermann:output-edge}
	}
	\If{$w_{n}\neq t$\label{alg:immermann:test-t2}}{\textbf{give up}}

	\caption{{\textsc Extension of the Immermann-Szelepscényi Theorem}}%
	\label{alg:immermann-szelepscenyi}
\end{algorithm}

We explain how to use the algorithm described in Lemma~\ref{lem:return-shortest-path-nl} to output a shortest path that satisfies some additional constraints.
\begin{lem}\label{lem:completion}
	Let $L \in \ttract$, let $(C,(v,q),e_K \cdots e_1)$ be an
	abbreviation and $E' \subseteq E$. There exists a nondeterministic logarithmic space algorithm that outputs a shortest
	trail $p$ such that $p \models_{E'} (C,(v,q),e_K \cdots e_1)$
	if it exists and rejects otherwise.
\end{lem}
\begin{proof}
	Let $G$ be a directed (labeled) multigraph.
	In order to find a path in $G$ that matches $C$, ends on $e_K \cdots e_1$, and
	uses edges $\{e_1,\dots,e_K\}$ only once, we use
	Algorithm~\ref{alg:immermann-szelepscenyi} on the product of $G$ (restricted
	to the edges $E' \cup \{e_1,\dots,e_K\}$) and $C$ extended with numbers $\ell
	\in [K]$. Since we cannot store the product in $O(\log n)$ space, we will construct it on-the-fly. Intuitively, the value of $\ell$ will tell us if we are in the last $K$ edges and if so, then which of the last $K$ edges we expect next.
	The ``product'' of $G$, $C$, and $[K]$ is a directed multigraph $G^*=(V^*,E^*,\edge^*)$ defined as follows:
	$V^* = V\times Q \times [K]$ and
	\begin{multline*}
		E^* = 	\{(e_\ell,(q_1,q_2,\ell))\mid \ell \in [K] \text{ and } \allowbreak  (q_1,\lab(e_\ell),q_2)\in C\}\\
		\cup \{(e,(q_1,q_2,K))\mid e \in E' -  \{e_1,\dots,e_K\} \text{ and } (q_1,\lab(e),q_2)\in C\}
	\end{multline*}
	\begin{align*}
		\edge^*((e,(q_1,q_2,K)))&= ((\orig(e),q_1,K), \lab(e), (\dest(e),q_2, K)) \text { if } e \neq e_K\\
		\edge^*((e,(q_1,q_2,K))) &= ((\orig(e),q_1,K), \lab(e), (\dest(e),q_2, K-1)) \text { if } e = e_K\\
		\edge^*((e_\ell,(q_1,q_2,\ell))) &= ((\orig(e_\ell),q_1,\ell), \lab(e_\ell), (\dest(e_\ell),q_2, \ell-1)) \text { if } \ell < K
	\end{align*}
	Since $K$ is a constant, the size of each state in $V^*$ is logarithmic in the input, and for two states $x,y \in V^*$, we can test in logarithmic space if there is an edge $e
	\in E^*$ such that $\edge^*(e)=(x,\lab(e),y)$. This is necessary for lines~\ref{alg:immermann:testedge1},~\ref{alg:immermann:testedge2}, and~\ref{alg:immermann:testedge3}.

	We then output a shortest path from $(v,q,K)$ to $(t,q',1)$ for $t$ being the target node of $e_1$ and some $q' \in C$.\footnote{Algorithm~\ref{alg:immermann-szelepscenyi} can also output a  shortest path from $s$ to some node in a set $T$ by testing $u \in T$ in line~\ref{alg:immermann:test-t} and $w_{n-1}\notin T$ in line~\ref{alg:immermann:test-t2}.} More precisely, since we want a path in $G$ and not in the product, we project away the unnecessary state and number and only output the corresponding edge in $G$ in line~\ref{alg:immermann:output-edge}.

	It remains to show that $p$ is a trail (in $G$).
	Assume towards contradiction that
	$p = d_1 \cdots d_m e_K \cdots e_1$ is not a trail. Then there exists
	an edge $d_i = d_j$ that appears at least twice in $p$. Note that
	$d_j$ is not in the suffix $e_K \cdots e_1$ by definition of $p$.
	We define
	\[
	p' = d_1 \cdots d_i d_{j+1} \cdots d_m e_K \cdots e_1
	\]
	and show that $p'$ is a shorter than $p$ but meets all requirements.
	Let $q_1 = \delta(q,d_1 \cdots d_i)$ and
	$q_2= \delta(q,d_1 \cdots d_j)$. By definition,
	$q_1, q_2 \in C$ and both have an incoming edge with label
	$\lab(d_i)=\lab(d_j)$. This allows us to use Corollary~\ref{cor:component}
	to ensure that
	\[\delta(q_1, d_{j+1} \cdots d_m e_K \cdots e_1)\in C.\] We can then apply Lemma~\ref{lem:sameState} to prove that
	\[
	\delta(q_1, d_{j+1} \cdots d_m e_K \cdots e_1) = \delta(q_2, d_{j+1} \cdots d_m e_K \cdots e_1) \; .
	\]
	So $p'$ is indeed a trail satisfying $p' \models_{E'}  (C,(v,q),e_K \cdots e_1)$. Furthermore,
	$p'$ is shorter than $p$, contradicting our assumption.
\end{proof}

Using the algorithm of Lemma~\ref{lem:completion} we can, in principle, output
a completion of $S$ that matches $L$ using nondeterministic logarithmic space.
However, such a completion does not necessarily correspond to a trail. The
reason is that, even though each trail $p_C$ we guess for some abbreviation
involving a component $C$
is a trail, the trails for different components may not be disjoint. Therefore, we will define pairwise
disjoint subsets of edges that can be used for the completion of the components.

The following definition fulfills the same purpose as the local
domains on nodes in Bagan et al.~\cite[Definition
7]{BaganBG-jcss20}. Since our components can be more complex, we
require extra conditions on the states (the $\delta_L(q,\pi) \in C$
condition).
\begin{defi}[Local Edge Domains]\label{def:locDomain}
	Let $S=\alpha_1 \cdots \alpha_k$ be a candidate summary and $E(S)$
	be the set of edges appearing in $S$. We define the local edge domains
	$\Edge_i \subseteq E_i$ inductively for each $i$ from $1$ to $k$,
	where $E_i$ are the remaining edges defined by  $E_1= E \setminus E(S)$
	and $E_{i+1}=E_i \setminus \Edge_i$.  If
	there is no trail $p$ such that $p \models \alpha_i$ or if
	$\alpha_i$ is a single edge, we define $\Edge_i =\emptyset$.

	Otherwise, let $\alpha_i=(C,(v,q),e_K \cdots e_1)$. We denote by $m_i$
	the minimal length of a trail $p$ with $p \models_{E_i} \alpha_i$ and
	define $\Edge_i$ as the set of edges used by trails $\pi$ that start at $v$, only
	use edges in $E_i$, are of length at most $m_i -K$, and satisfy
	$\delta_L(q,\pi) \in C$.
\end{defi}

By definition of $\Edge_i$, we can conclude that $\edge(e_i) \neq \edge(e_j)$
for all $e_i \in \Edge_i, e_j \in \Edge_j, i \neq j$, as $e_i \in \Edge_i$ and
$\edge(e_i)=\edge(e_j)$ imply that $e_j \in \Edge_i$. We note that a shortest
trail using $e_i$ but not $e_j$ can use $e_j$ instead of $e_i$.
We note that %
the sets $E(S)$ and $(\Edge_i)_{i\in[k]}$ are always disjoint.

\begin{defi}[Admissible Trail]
  We say that a trail $p$ is \emph{admissible} if there exist a
  candidate summary $S=\alpha_1 \cdots \alpha_k$ and trails
  $p_1,\dots,p_k$ such that $p=p_1 \cdots p_k$ is a completion of $S$
  and $p_i \models_{\Edge_i} \alpha_i$ for every $i\in [k]$.
\end{defi}

\begin{figure}[t] \centering
    \begin{minipage}{.3\textwidth}
    \begin{tikzpicture}[auto,>=latex]
    \node [label={[xshift=0.05cm, yshift=-.1cm]$v$}] (q2) at (-2.3,0) {};
    \node (q3) at (0,0) {};
    \node (q5) at (-.9,-.45) {};
    \node(q6) at (-1.5,-.7) {};
    \node(qs) at (-3.5,0) {};
    \node(qt) at (-2.6,-.9) {};

        \path[draw,<-,decorate,
    decoration={snake,amplitude=.3mm,segment length=2.5mm,pre=lineto,pre
        length=5pt}] (q2) -- (qs);
        \path[draw,<-,decorate,
    decoration={snake,amplitude=.3mm,segment length=2.5mm,pre=lineto,pre
        length=5pt}] (qt) -- (q6);

      \foreach \i in {2,3,5,6}{
           \fill (q\i) circle (2pt);
       }
    \path (q5) edge [->] node {$e$} (q6);
    \path (q2) edge [dotted, ->,swap] node {$\pi$} (q5);

    \path[draw,<-,decorate,
    decoration={snake,amplitude=.3mm,segment length=2.5mm,pre=lineto,pre
        length=5pt}] (q3) -- (q2);

     \node(elli) at ($(q2)!0.5!(q3)$) {};
    \draw (elli) ellipse (1.3cm and .9cm);
    \node(edgei) at (elli |- {{(0,1.2)}})  {$\Edge_i \cup \{e_1,\dots e_K\}$};

     \path(q5) edge [ in =-20, out = -20,draw,<-,decorate,
    decoration={snake,amplitude=.3mm,segment length=2.5mm,pre=lineto,pre
        length=5pt}]  (q3);
    \end{tikzpicture}
    \end{minipage}
\hspace{.1cm}
    \begin{minipage}{.5\textwidth}
    \begin{tikzpicture}[auto,>=latex]
    \node [label={[xshift=0.05cm, yshift=-.1cm]$v$}]  (q2) at (-2.3,0) {};
    \node (q3) at (2,0) {};
    \node(v) at (0.2,0) {};
    \node (q5) at (-1.44,-1.1) {};
    \node(q6) at (-.8,-1.1) {};
    \node(qs) at (-3.5,0) {};
    \node(qt) at (3.5,0) {};

    \path[draw,<-,decorate,
    decoration={snake,amplitude=.3mm,segment length=2.5mm,pre=lineto,pre
        length=5pt}] (q2) -- (qs);
    \path[draw,<-,decorate,
    decoration={snake,amplitude=.3mm,segment length=2.5mm,pre=lineto,pre
        length=5pt}] (qt) -- (q3);

    \foreach \i in {2,3,5,6}{
        \fill (q\i) circle (2pt);
    }
    \path (v) edge [->,swap] node {$e_K \cdots e_1$} (q3);
    \path (q5) edge [->,swap] node {$e$} (q6);
    \path (q2) edge [dotted, ->] node {} (q3);
    \draw [decorate,decoration={brace,amplitude=10pt},xshift=-4pt,yshift=0pt]
    (q2) -- (q3) node [black,midway,xshift=-0cm,yshift = .3cm] {$\pi$};

    \node(elli) at ($(q2)!0.5!(q3)$) {};
    \draw (elli) ellipse (2.4cm and .8cm);
    \node(edgei) at (elli |- {{(0,1.1)}})  {$\Edge_\ell \cup \{e_1,\dots,e_K\}$};

    \path(q5) edge [ draw,<-,decorate,
    decoration={snake,amplitude=.3mm,segment length=2.5mm,pre=lineto,pre
        length=5pt}]  (q2);

    \path (q3 -| {{(.3,0)}}) edge [ draw,<-,decorate,
    decoration={snake,amplitude=.3mm,segment length=2.5mm,pre=lineto,pre
        length=5pt}]  (q6);
    \end{tikzpicture}
\end{minipage}
    \caption{Sketch of case (1) and (2) in the proof of Lemma~\ref{lem:Bagan13}}%
    \label{fig:Bagan13}
\end{figure}

We show that \emph{shortest}
trails that match $L$ are always admissible. Thus, the existence of a trail is
equivalent to the existence of an admissible trail.
\begin{lem}\label{lem:Bagan13}
  Let $G$ and $(s, t)$ be an instance for $\rtq(L)$, with $L \in \ttract$.
  Then every shortest trail from $s$ to $t$ in $G$ that matches $L$ is admissible.
\end{lem}
\begin{proof}[Proof sketch]
  We assume towards a contradiction that there is a shortest trail $p$ from $s$ to $t$ in
  $G$ that matches $L$ and is not admissible. That means there is some $\ell \in
  \nat$, and an edge $e$ used in $p_\ell$ with $e \notin \Edge_\ell$. There are
  two possible cases: (1) $e \in \Edge_i$ for some $i < \ell$ and (2) $e \notin
  \Edge_i$ for any $i$. In both cases, we construct a shorter trail $p$ that
  matches $L$, which then leads to a contradiction. We depict the two cases in
  Figure~\ref{fig:Bagan13}. We construct the new trail by substituting the
  respective subtrail with $\pi$.
\end{proof}
\begin{proof}
  \newcommand{\ropen}[1]{[#1)} 
  \newcommand{\lopen}[1]{(#1]} 
  In this proof, we use the following notation for trails. By $p\ropen{e_1,e_2}$ we denote the prefix of $p[e_1,e_2]$ that excludes the last edge (so it can be empty). Notice that $p[e_1,e_2]$ and $p\ropen{e_1,e_2}$ are always well-defined for trails.
  Let $p=d_1 \cdots d_m$ be a shortest trail from $s$ to $t$ that
  matches $L$. Let $S=\alpha_1 \cdots \alpha_k$ be the summary of $p$ and let $p_1,\dots,p_k$ be trails such that $p=p_1 \cdots p_k$ and
  $p_i \models \alpha_i$ for all $i \in [k]$. We denote by $\Left_i$ and
  $\Right_i$ the first and last edge in $p_i$. By definition of $p_i$ and the
  definition of summaries, $\Left_i$ and $\Right_i$ are identical with $\Left_C$
  and $\Right_C$ if $\alpha_i \in \Abbrv$ is an abbreviation for the component
  $C$.

  Assume that $p$ is not admissible.
  That means there is some edge $e$
  used in $p_\ell$ such that $e \notin \Edge_\ell$. There are two possible
  cases:
  \begin{enumerate}[(1)]
  \item $e \in \Edge_i$ for some $i < \ell$; and
  \item $e \notin \Edge_i$ for any $i$.
  \end{enumerate}
  In case (1), we choose $i$ minimal such that some edge
  $e \in \Edge_i$ is used in $p_j$ for some $j>i$. Among all such
  edges $e \in \Edge_i$, we choose the edge that occurs latest in $p$.
  This implicitly maximizes $j$ for a fixed $i$. Especially no edge
  from $\Edge_i$ is used in $p_{j+1} \cdots p_k$.

  Let $\alpha_i=(C_i,(v,q),e_K \cdots e_1)$. By definition of
  $\Edge_i$, there is a trail $\pi$ from $v$, ending with $e$, with $\delta_L(q,\lab(\pi)) \in C_i$,
  and that is shorter than the subpath
  $p[\Left_i,\Right_i]$ and therefore shorter than $p[\Left_i,e]$.

  We now show that $p' = p_1 \cdots p_{i-1} \pi p\lopen{e,d_m}$
  is a trail. Since $p$ is a trail, it suffices to prove that the edges
  in $\pi$ are disjoint with other edges in $p'$. We note that all
  intermediate edges of $\pi$ belong to $\Edge_i$. By minimality of
  $i$, no edge in $p_1 \cdots p_{i-1}$ can use any edge of $\Edge_i$
  and by our choice of $e$, no edge in $p$ after $e$ can use any edge of
  $\Edge_i$. This shows that $p'$ is a trail.

  We now show that $p'$ matches $L$. Since
  $e$ appears in $p_j$, there is a path from $\Left_j$ to $\Right_j$
  over $e$ that stays in $C_j$. Let $q_1$ and $q_2$ be the states of
  $A_L$ before and after reading $e$ in $p$ and, analogously, $q_1'$
  and $q_2'$ the states of $A_L$ before and after reading $e$ in
  $p'$. That is
  \begin{align*}
    q_1 \quad&=\quad \delta_L(i_L,p\ropen{d_1,e}) & q_2\quad&=\quad \delta_L(q_1,e) \\
    q_1' \quad&=\quad \delta_L(i_L,p'\ropen{d_1,e}) & q_2'\quad&=\quad \delta_L(q_1',e)
  \end{align*}
  We note that in $p'$, $e$ is at the end of the subtrail $\pi$.

  We can conclude that the states $q_1$ and $q_1'$ both have loops
  starting with $a=\lab(e)$, as the transition $(q_1,\lab(e),q_2)$ is read in
  $C_j$ and the transition $(q_1',\lab(e),q_2')$ is read in $C_i$.
  Furthermore, $q'_1 \leadsto q_1$, since $q'_1 \in C_i$ and $q_1
  \in C_j$. Therefore, Lemma~\ref{lem:Bagan9} implies that $\cL_{q_1} \cap
  L_{q_1}^{a}\Sigma^* \subseteq \cL_{q'_1}$ where $L_{q_1}^{a}$ denotes all words $w$ of length $K$ that start
  with $a$ and such that $\delta_L(q_1, w) \in C_j$.

  We have that $\lab(p[e,d_m]) \in \cL_{q_1}$ by the fact that $p$
  matches $L$. We have that $\lab(p[e,d_m]) \in \cL_{q_1}^a\Sigma^*$, as, by
  the definition of summaries, $A_L$ stays in $C_j$ for at least $K$
  more edges after reading $e$ in $p$. We can conclude that
  $\lab(p[e,d_m]) \in \cL_{q_1'}$, which proves that $p'$ matches
  $L$.

  This concludes case (1). For case (2), we additionally assume w.l.o.g.\ that there is no
  edge $e \in \Edge_i$ that appears in some $p_j$ with $j>i$, i.e., no edge satisfies case (1).
  By definition of $\Edge_\ell$, there is a trail $\pi$ with
  $\pi \models_{\Edge_\ell} \alpha_\ell$ that is shorter than
  $p[\Left_\ell,\Right_\ell]$.
  We choose $p'$ as the path obtained from $p$ by replacing $p_\ell$
  with $\pi $.

  We now show that
  $p' = p_1  \cdots p_{\ell-1} \cdot \pi \cdot p_{\ell+1} \cdots p_k$ is a
  trail. Since $p$ is a trail, it suffices to prove that the edges in
  $\pi $ are disjoint with other edges in $p'$. We note that all
  intermediate edges of $\pi$ belong to $\Edge_\ell$.

  By definition of $\Edge_\ell$, no edge in
  $p_1 \cdots p_{\ell-1}$ is in $\Edge_\ell$. And by the assumption that
  there is no edge satisfying case (1), %
  no edge in $p_{\ell+1} \cdots p_k$ is in $\Edge_\ell$. Therefore, $p'$ is  a trail.

  It remains to prove that $p'$ matches $L$. Let $(C,(v,\hat{q}),e_K \cdots e_1) = \alpha_\ell$ and let $q$ and $q'$ be the
  states in which $A_L$ is before reading $e_{K}$ in $p$ and $p'$,
  respectively. By definition of a summary, we have that
  $\delta_L(q,e_K \cdots e_1) \in C$ and, by definition of
  $\models$, we have that $\delta_L(q',e_K \cdots e_1) \in C$.
  By Lemma~\ref{lem:sameState} we can conclude that
  $\delta_L(q,e_K \cdots e_1)=\delta_L(q',e_K \cdots e_1)$.
  As $p$ matches $L$, we can conclude that also $p'$ matches $L$.
\end{proof}

So, if there is a solution to $\rtq(L)$, we can find it by enumerating the candidate summaries and
completing them using the local edge domains.
We next prove that testing if an edge is in $\Edge_i$ can be done in
logarithmic space.
We will name this decision problem $\pedge(L)$ and define it as follows:
\decisionproblem{.76\linewidth}{$\pedge(L)$}{A (multi-)graph $G = (V,E,\edge)$, nodes $s, t$, a 
  candidate summary $S$, an edge $e \in E$ and an integer $i$.}{Is $e \in
  \Edge_i$?}

\begin{lem}\label{lem:computeEdgei}
  $\pedge(L)$ is in NL for every $L \in \ttract$.
\end{lem}
\begin{proof}
    The proof is similar to the proof of Lemma~17 by Bagan et al.~\cite{BaganBG-jcss20}, which
    is based on the following result due to Immerman~\cite{Immerman-siamcomp88}:
    $\text{NL}^{\text{NL}}=\text{NL}$. In other words, if a decision problem $P$
    can be solved by an NL algorithm using an oracle in NL, then this problem
    $P$ belongs to NL\@. Let, for each $k\geq 0$, $\pedge^{\leq k}(L)$ be the
    decision problem $\pedge(L)$ with the restriction $i\leq k$, i.e.,
    $(G,s,t,S,e,i)$ is a positive instance of $\pedge^{\leq k}(L)$ if and only if
    $(G,s,t,S,e,i)$ is a positive instance of $\pedge(L)$ and $i\leq k$. Notice
    that $i$ belongs to the input of $\pedge^{\leq k}(L)$ while this is not the
    case for $k$. Obviously, $\pedge(L)=\pedge^{\leq |S|}(L)$.
    We prove that $\pedge^{\leq k}(L) \in \text{NL}$ for each $k\geq 0$. If $k=0$,
    $\pedge^{\leq 0}(L)$ always returns False because $\Edge_i$ is not defined for
    $i=0$. So $\pedge^{\leq 0}(L)$ is trivially in NL\@. Assume, by induction, that
    $\pedge^{\leq k}(L) \in \text{NL}$. It suffices to show that there is an
    NL algorithm for $\pedge^{\leq k+1}(L)$ using $\pedge^{\leq k}(L)$ as an
    oracle. Since $\text{NL}^{\text{NL}}=\text{NL}$, this implies that
    $\pedge^{\leq k+1}(L)\in \text{NL}$.

    Let $(G,s,t,S,e,i)$ be an instance of $\pedge^{\leq k+1}(L)$. If
    $i\leq k$, we return the same answer as the oracle
    $\pedge^{\leq k}(L)$. If $i=k+1$ and $\alpha_i \in E$, we return False, as $\Edge_i= \emptyset$.
    If $i=k+1$ and $\alpha_i \in \Abbrv$, we first compute the length $m$
    of a minimal trail $p$ such that $p \models_{E_i} \alpha_i$ using
    the NL algorithm of Lemma~\ref{lem:completion}. We note that we can compute
    $E_i$ using the NL algorithm for $\pedge^{\leq k}$.

    To test whether the edge $e$ can be used by a trail from some
    $(v,q)$ in at most $m-K$ steps, we use the
    on-the-fly product of $G$ and $A_L$ restricted to the
    edges in $E_i$ and states in $C$. We search for a shortest
    path
    from
    $(v,q)$ to some $(v',q') \in V \times C$ that ends with $e$. We
    recall that reachability is in \nlogspace.

    We note that this trail in the product graph might correspond to a
    path $p$ with a cycle in $G$. As we project away the states, some
    distinct edges in the product graph are actually the same edge in
    $G$. However, by Lemma~\ref{lem:component}, we can remove all
    cycles from $p$ without losing the property that
    $\delta_L(q,p) \in C$.  This concludes the proof.
\end{proof}

With this, we can finally give an NL algorithm that decides whether
a candidate summary can be completed to an admissible trail that matches $L$.

\begin{lem}\label{lem:algo} %
  Let $L$ be a language in \etract. There exists an NL
  algorithm that given an instance $G$, $(s,t)$ of $\rtq(L)$ and a candidate
  summary $S=\alpha_1 \cdots \alpha_k$ tests whether there is a trail $p$ from $s$
  to $t$ in $G$ with summary $S$ that matches $L$.
\end{lem}
\begin{proof}
  We propose the following algorithm, which consists of three tests:
  \begin{enumerate}[(1)]
  \item Guess, on-the-fly, a path $p$ from $S$ by replacing each $\alpha_i$ by a
    trail $p_i$ such that $p_i \models_{\Edge_i} \alpha_i$ for each $i \in [k]$.
    This test succeeds if and only if this is possible.
  \item In parallel, check that $p$ matches $L$.
  \item In parallel, check that $S$ is a summary of $p$.
  \end{enumerate}
  We first prove that the algorithm is correct. First, we assume that there is a
  trail with summary $S$ from $s$ to $t$ that matches $L$. Then, there is also a
  shortest such trail and, by Lemma~\ref{lem:Bagan13}, this trail is admissible.
  Therefore, the algorithm will succeed.

  Conversely, assume that the algorithm succeeds. Since $E(S)$ and all the sets
  $\Edge_i$ are mutually disjoint, the path $p$ is always a trail. %
  By tests (2) and (3), it is a trail from $s$ to $t$ that matches $L$.

  We still have to check the complexity. We note that the sets $\Edge_i$ are not
  stored in memory: we only need to check on-the-fly if a given edge belongs to
  those sets, which only requires logarithmic space according to
  Lemma~\ref{lem:computeEdgei}. Therefore, we use an on-the-fly adaption of the
  NL algorithm from Lemma~\ref{lem:completion}, which requires a set $\Edge_i$
  as input, which we will provide on-the-fly.

  Testing if $p$ matches $L$ can simply be done in parallel to test (1) on an
  edge-by-edge basis, by maintaining the current state of $A_L$ in memory. If we
  do so, we can also check in parallel if $S = \alpha_1 \cdots \alpha_k$ is a
  summary of $p$. This is simply done by checking, for each $\alpha_i$ of the
  form $(C,(v,q),e_K \cdots e_1)$ and $\alpha_{i+1} = e$, whether $e \notin C$.
  This ensures that, after being in $C$ for at least $K$ edges, the path $p$
  leaves the component $C$, which is needed for summaries. Furthermore, we test
  if there is no substring $\alpha_i \cdots \alpha_j$ in $S$ that purely
  consists of edges that are visited in the same component $C$, but which is too
  long to fulfill the definition of a summary. Since this maximal length is a
  constant, we can check it in \nlogspace.
\end{proof}

We eventually show the main Lemma of this section, proving that $\rtq(L)$ is
tractable for every language in \etract.
\begin{lem}\label{lem:upperBound} Let $L \in \etract$. Then, $\rtq(L)$ $\in$
  NL\@.
\end{lem}
\begin{proof} We simply enumerate all possible candidate summaries $S$ w.r.t.
  $(L,G,s,t)$, and apply on each summary the algorithm of Lemma~\ref{lem:algo}. We
  return \true if this algorithm succeeds and \false otherwise. Since the
  algorithm succeeds if and only if there exists an admissible path from
  $s$ to $t$ that matches $L$, and consequently, if and only if there is a trail from $s$ to $t$ that matches $L$ (Lemma~\ref{lem:Bagan13}), this is the right answer.
  Since $L$ is fixed, there is a polynomial number of candidate summaries, each of
  logarithmic size. Consequently, they can be enumerated within logarithmic space.
\end{proof}

\begin{lem}
  Let $L \in \etract$ and $L$ be infinite. Then, $\rtq(L)$ is NL-complete.
\end{lem}
\begin{proof}
  The upper bound is due to Lemma~\ref{lem:upperBound}, the lower due to
  reachability in directed graphs being NL-hard.
\end{proof}

\begin{cor}\label{cor:output-shortest-paths}
  Let $L \in \ttract$, $G$ be a multigraph, and $s$, $t$ be nodes in $G$. If there
  exists a trail from $s$ to $t$ that matches $L$, then we can output a shortest
  such trail in polynomial time (and in nondeterministic logarithmic space).
\end{cor}
\begin{proof}
  For each candidate summary $S$, we first use Lemma~\ref{lem:algo} to decide
  whether there exists an admissible trail with summary $S$. With the algorithm
  in Lemma~\ref{lem:completion}, we then compute the minimal length $m_i$ of
  each $p_i$. The sum of these $m_i$s then is the length of a shortest trail
  that is a completion of $S$. We will keep track of a summary of one of the
  shortest trails and finally recompute the overall shortest trail completing
  this summary and outputting it. Notice that this algorithm is still in NL
  since the summaries have constant size and overall counters never exceed $|E|$.
\end{proof}

\subsection{Languages not in \texorpdfstring{\ttract}{Ttract}}
In this section we prove that $\rtq(L)$ is NP-hard for languages $L \notin \ttract$, even if the input is restricted to graphs. Therefore, NP-completeness also follows for multigraphs.
The proof of Theorem~\ref{theo:trichotomy}\ref{theo:trichotomy-np}
is by reduction from the following \np-complete problem:
\decisionproblem{.75\linewidth}{$\edgedispaths$}{A language $L$, a graph $G = (V,E,\edge)$,
  and two pairs of nodes $(s_1,t_1)$, $(s_2,t_2)$.}{Are there two paths $p_1$ from
  $s_1$ to $t_1$ and $p_2$ from $s_2$ to $t_2$ such that $p_1$ and $p_2$ are
  edge-disjoint?}
The proof is very close to the corresponding proof for simple paths
by Bagan et al.~\cite[Lemma~4]{BaganBG-jcss20} (which is a reduction from the
two vertex-disjoint paths problem).

\begin{lem}
  $\edgedispaths$ is NP-complete.
\end{lem}
\begin{proof}
  Fortune et al.~\cite{FortuneHW-TCS80} showed that the problem variant of
  \edgedispaths that asks for node-disjoint paths is NP-complete. The reductions
  from LaPaugh and Rivest~\cite[Lemma 1 and 2]{LapaughR-jcss80} or Perl and Shiloach~\cite[Theorem 2.1 and 2.2]{PerlS-jacm78} then imply that
  the NP completeness also holds for edge-disjoint paths. %
\end{proof}

To prove the
lower bound, we first show that every regular language that is not in \etract
admits a witness for hardness, which is defined as follows.
\begin{defi} A \emph{witness for hardness} is a tuple $(q,w_m, w_r,w_1,
  w_2)$ with $q\in Q_L$, $w_m,w_r, w_1,w_2 \in \Sigma^*$, $w_1 \in \loopwords(q)$ and there exists
  a symbol $a\in \Sigma$ with $w_1 = a w'_1$ and $w_2 = a w'_2$ and satisfying
  \begin{itemize}
  \item $w_m (w_2)^* w_r \subseteq \cL_{q}$, and
  \item $(w_1+w_2)^*w_r \cap \cL_{q} = \emptyset$.
  \end{itemize}
\end{defi}

\noindent
Before we prove that each regular language that is not in \etract has such a
witness, recall Property $P$:
\[ \cL_{q_2}\subseteq \cL_{q_1} \text{ for all } q_1, q_2 \in Q_L \text{ such that }
q_1 \leadsto q_2 \text{ and } \loopwords(q_1)\cap
\loopwords(q_2)\neq \emptyset\]

\begin{lem}\label{lem:witness}
  Let $L$ be a regular language that does not
  belong to \etract. Then, $L$ admits a witness for hardness.
\end{lem}
\begin{proof}
   Let $L$ be a regular language that does not belong to $\etract$.
  Then there exist $q_1,q_2 \in Q_L$ and words
  $w_1, w_2$ with $w_1 = a w'_1$ and $w_2 = a w'_2$ such that $w_1 \in
  \loopwords(q_1), w_2 \in \loopwords(q_2)$, and $q_1 \leadsto q_2$ such that
  $w_2^M w'_r \notin \cL_{q_1}$ for a $w'_r \in \cL_{q_2}$. Let $w_m$ be a word with $q_2
  = \delta_L(q_1, w_m)$. We set $w_r = w_2^M w'_r$.

  We now show that the so-defined tuple $(q_1,w_m, w_r,w_1, w_2)$ is a witness for
  hardness. By definition, we have $w_m (w_2)^*w_r
  \subseteq \cL_{q_1}$. We distinguish two cases, depending on whether
  $L$ satisfies Property $P$ or not. %
If $L$ does not satisfy $P$,
we can assume \mbox{w.l.o.g.} that in our tuple we have $w_1=w_2$ and since $w_2^M
w'_r \notin \cL_{q_1}$, we also have $w_2^* w_r \cap \cL_{q_1} \neq \emptyset$, so it is indeed a witness for hardness.

Otherwise, $L$ is aperiodic, see Lemma~\ref{lem:p}. We assume
\mbox{w.l.o.g.} that $w_1=({w'_1})^M$ for some word ${w'_1}$. Then, we claim that $\cL_{q'}
\subseteq \cL_{q_1}$ for every $q'$ in $\delta_L(q_1,\Sigma^*w_1)$. %
Indeed, every $q'\in \delta_L(q_1,\Sigma^*w_1)$ loops over $w_1$ by the pumping
lemma and aperiodicity of $L$, hence $w_1 \in \loopwords(q_1)\cap \loopwords(q')$ and
therefore $\cL_{q'} \subseteq \cL_{q_1}$ due to Property $P$.

It remains to  prove that $(w_1+w_2)^*w_r \cap \cL_{q_1} = \emptyset$. Every word in
$(w_1+w_2)^*w_r$ can be decomposed into $uv$ with $u \in \varepsilon+
(w_1+w_2)^*w_1$ and $v \in w_2^*w_r$. For $q' = \delta_L(q_1,u)$ we have
proved that $\cL_{q'} \subseteq \cL_{q_1}$, so it suffices to show that $v \notin
\cL_{q_1}$. This is immediate from $w_r = w_2^M w'_r \notin \cL_{q_1}$ and
the aperiodicity of $L$. So we have $uv \notin\cL_{q_1}$ and the tuple $(q,w_m,
w_r,w_1, w_2)$ is indeed a witness for hardness.
\end{proof}

We can now show the following
\begin{lem}\label{lem:Bagan2}
  Let $L$ be a regular language that does not belong to \etract. Then, $\rtq(L)$
  is NP-complete.
\end{lem}

\newcommand\xrsquigarrow[1]{%
\mathrel{\begin{tikzpicture}[baseline= {( $ (current bounding box.south) + (0,0ex) $ )}]
\node[inner sep=.5ex] (n) {$\scriptstyle \smash[t]{#1}$};
\path[draw,<-,decorate,
  decoration={snake,amplitude=0.5pt,segment length=1.2mm,pre=lineto, pre length=2pt}]
    (n.south east) -- (n.south west);
\end{tikzpicture}
}%
}

\newcommand{\mypath}[3]{#1 \xrsquigarrow{#2} #3\xspace}

\begin{proof}
  The proof is almost identical to the reduction from two node-disjoint paths to
  the $\rspq(L)$ problem by Bagan et al.~\cite{BaganBG-jcss20}. Clearly,
  $\rtq(L)$ is in NP for every regular language $L$, since we only need to guess
  a trail of length at most $|E|$ from $s$ to $t$ and verify that the word on
  the trail is in $L$. Let $L \notin \etract$. We exhibit a reduction from
  $\edgedispaths$ to $\rtq(L)$. According to Lemma~\ref{lem:witness}, $L$ admits
  a witness for hardness $(q,w_m, w_r,w_1, w_2)$. Let $w_\ell$ be a word such
  that $\delta_L(i_L,w_\ell) = q$. By definition of a witness we get
  $w_\ell (w_1 + w_2)^* w_r \cap L = \emptyset$ and $w_\ell w_1^* w_m w_2^* w_r
  \subseteq L$. Let $a \in \Sigma$ and $w'_1,w'_2 \in \Sigma^*$ such that $w_1 =
  a w'_1$ and $w_2 = a w'_2$. If $w'_1$ or $w'_2$ is empty, we replace it with
  $a$.

  In the following construction, whenever we say that we add a path from $v_0$
  to $v_n$ labeled by a word $w=a_1 \cdots a_n$, denoted by $\mypath{v_0}{w}{v_n}$
  we mean that we add $n-1$ new nodes $v_1,\dots,v_{n-1}$ and $n$ new edges
  $e_1,\dots,e_n$ such that $\edge'(e_i)=(v_{i-1},a_i,v_i)$.

  Let $G=(V,E)$ be an unlabeled and simple input graph for the $\edgedispaths$
  problem and $s_1,t_1,s_2,t_2$ be nodes in $V$. We build from $G$ a graph
  $G'=(V',E',\edge')$ such that $(G,s_1,t_1,s_2,t_2)$ is a yes-instance of
  \edgedispaths if and only if there is a trail from $s$ to $t$ matching $L$ in $G'$.
  We start with the nodes from $G$ and add two new nodes $s$ and $t$ and three paths
  $\mypath{s}{w_\ell}{s_1}$, $\mypath{t_1}{w_m}{s_2}$, and $\mypath{t_2}{w_r}{t}$.
  Furthermore, for each edge $(v_1,v_2)$ in $G$, we add a new node $v_{12}$ and
  three paths $\mypath{v_1}{a}{v_{12}}$, $\mypath{v_{12}}{w'_1}{v_2}$, and
  $\mypath{v_{12}}{w'_2}{v_2}$.
  An example for the language $da^*c(abc)^*ef$ and some graph $G$ can be seen in Figure~\ref{fig:full-lower}.

  By construction, two edge disjoint paths $p_1$ and $p_2$ in $G$ going from
  $s_1$ to $t_1$ and from $s_2$ to $t_2$ correspond to a trail $p$ from $s$ to $t$
  in $G'$ that contains the path $\mypath{t_1}{w_m}{s_2}$. Such a trail $p$
  matches a word in $w_\ell (w_1+w_2)^* w_m (w_1+w_2)^* w_r$. And, as paths labeled $w_1$ and $w_2$ can be used interchangeably, we find a trail matching $w_\ell w_1^* w_m w_2^* w_r \subseteq L$.

  For the other direction, we have to show two things. First, we show that every
  trail $p$ in $G'$ from $s$ to $t$ that uses the path $\mypath{t_1}{w_m}{s_2}$
  proves the existence of two edge disjoint paths $p_1$ and $p_2$ in $G$ from
  $s_1$ to $t_1$ and from $s_2$ to $t_2$. Indeed $p_1$ and $p_2$ can be computed
  from $p$ by keeping only those nodes that are from $G$ and splitting $p$
  between $t_1$ and $s_2$. The paths are disjoint, as otherwise some edge $v_i$
  to $v_{ij}$ has to be used twice by $p$. Second, we show that there can be no
  trail $p$ from $s$ to $t$ in $G'$ that matches $L$ and does not use the path
  $\mypath{t_1}{w_m}{s_2}$. Indeed, every trail $p$ from $s$ to $t$ in $G'$ that
  does not contain the path $\mypath{t_1}{w_m}{s_2}$ matches a word in $w_\ell
  (w_1+w_2)^*w_r$. By definition of witness for hardness, no such word is in
  $L$. Thus, $\rtq(L)$ returns \true for $(G',s,t)$ if and only if there is a
  trail from $s$ to $t$ in $G'$ that contains the edge $(t_1, w_m, s_2)$ that
  is, if and only if \edgedispaths returns \true for $(G,s_1,t_1,s_2,t_2)$.
\end{proof}

\begin{figure}[t]
  \hfill
		\begin{tikzpicture}[auto,>=latex,baseline=0pt]%
		\draw (-.5,.4) coordinate[label={left:$G$}];
			\node (q1) at (.4,0) {};
			\node (q2) at (2,0) {};
			\node (q3) at (4-.4,0) {};

			\node (q4) at (.4,-2) {};
			\node (q5) at (2,-2) {};
			\node (q6) at (4-.4,-2) {};

			\foreach \i in {1,2,3,4,5,6}{
				\fill (q\i) circle (2pt);
			}

			\draw (q1) coordinate[label={left:$s_1$}];
			\draw (q3) coordinate[label={right:$t_1$}];
			\draw (q6) coordinate[label={right:$s_2$}];
			\draw (q4) coordinate[label={left:$t_2$}];

			\path (q1) edge [->] node {} (q2);
			\path (q2) edge [->] node {} (q3);

			\path (q6) edge [->] node {} (q5);
			\path (q5) edge [->] node {} (q4);
			\path (q2) edge [->] node {} (q5);
		\end{tikzpicture}
    \hfill
		\begin{tikzpicture}[auto,>=latex,baseline=0pt]%
		\draw (-1.5,.4) coordinate[label={left:$G'$}];
		\node (s) at (-1,0) {};
		\node (t) at (-1,-2) {};

		\node (q1) at (0,0) {};
		\node (q2) at (2,0) {};
		\node (q3) at (4,0) {};
		\node (q12) at (1,0) {};
		\node (q23) at (3,0) {};

		\node (q4) at (0,-2) {};
		\node (q5) at (2,-2) {};
		\node (q6) at (4,-2) {};
		\node (q45) at (1,-2) {};
		\node (q56) at (3,-2) {};

		\node (q25) at (2,-.9) {};

		\foreach \i in {1,2,3,4,5,6,12,23,45,56,25}{
			\fill (q\i) circle (2pt);
		}

		\draw (q1) coordinate[label={above:$s_1$}];
		\draw (q3) coordinate[label={right:$t_1$}];
		\draw (q6) coordinate[label={right:$s_2$}];
		\draw (q4) coordinate[label={below:$t_2$}];
		\draw (s) coordinate[label={left:$s$}];
		\draw (t) coordinate[label={left:$t$}];

		\path (q1) edge [->] node {$a$} (q12);
		\path (q12) edge [->, bend left = 15] node {$a$} (q2);
		\path (q12) edge [->, bend right = 15,swap ] node {$bc$} (q2);

		\path (q2) edge [->] node {$a$} (q23);
		\path (q23) edge [->, bend left = 15] node {$a$} (q3);
		\path (q23) edge [->, bend right = 15,swap ] node {$bc$} (q3);

		\path (q2) edge [->] node {$a$} (q25);
		\path (q25) edge [->, bend left = 15] node[pos=0.45] {$a$} (q5);
		\path (q25) edge [->, bend right = 15,swap ] node[pos=0.45] {$\smash[t]{b}c$} (q5);

		\path (q5) edge [->] node {$a$} (q45);
		\path (q45) edge [->, bend left = 15] node {$bc$} (q4);
		\path (q45) edge [->, bend right = 15,swap ] node {$a$} (q4);

		\path (q6) edge [->] node {$a$} (q56);
		\path (q56) edge [->, bend left = 15] node {$bc$} (q5);
		\path (q56) edge [->, bend right = 15,swap ] node {$a$} (q5);

		\path (q3) edge [->,bend left = 15] node {$c$} (q6);

		\fill (s) circle (2pt);
		\fill (t) circle (2pt);

		\path (s) edge [->] node {$d$} (q1);
		\path (q4) edge [->] node {$ef$} (t);
  \end{tikzpicture}
  \hfill\mbox{}
	\caption{Example of the reduction in Lemma~\ref{lem:Bagan2} for the language
    $d a^* c (abc)^* ef$. We use $w_\ell=d$, $w_m=c$, $w_r=ef$, $w_1=aa$, and
    $w_2=abc$ for the construction. For the ease of readability, we omit the
    intermediate nodes on the $bc$ and $ef$ paths.}%
	\label{fig:full-lower}
\end{figure}

\section{Recognition and Closure Properties}\label{sec:closureproperties}

The following theorem establishes the complexity of deciding
if a regular language is in \ttract.

  Before we establish the complexity of deciding for a regular
  language $L$ whether $L \in \ttract$, we need some lemmas. The first has been adapted from the simple path case (Lemma~6
  in~\cite{BaganBG-jcss20}).
\begin{lem}\label{lem:Bagan4}
    Let $L$ be a regular language. Then, $L$ belongs to \ttract if and only if for all pairs of states $q_1, q_2 \in Q_L$ and symbols $a \in \Sigma$ such that $q_1 \leadsto q_2$ and $\loopwords(q_1)\cap a\Sigma^* \neq \emptyset$, the following statement holds: $(\loopwords(q_2) \cap a\Sigma^*)^N\cL_{q_2}\subseteq \cL_{q_1}$.
\end{lem}
\begin{proof}
    The (if) implication is immediate by Corollary~\ref{cor:leftsync}.
    Let us now prove the (only if) implication.
    Since the proof of this lemma requires a number of different states and words, we
    provide a sketch in Figure~\ref{fig:Bagan4}.
     Assume $L \in \ttract$.
    Let $q_1,q_2$ be two states such that $\loopwords(q_1)\cap a\Sigma^* \neq \emptyset$ and $q_1 \leadsto q_2$.
    If $\loopwords(q_2) \cap a \Sigma^* = \emptyset$, the statement follows immediately. So let us assume w.l.o.g.\ that $\loopwords(q_2) \cap a \Sigma^*\neq \emptyset$.
    Let $v_1, \ldots, v_N \in (\loopwords(q_2) \cap a \Sigma^*)$ be arbitrary words and $q_3=\delta_L(q_1,v_1\cdots v_N)$.
    We want to prove $\cL_{q_2}\subseteq \cL_{q_3}$.
    For some $i,j$ with $0 \leq i < j \leq N$, we get $\delta_L(q_1,v_1\cdots v_i)=\delta_L(q_1,v_1 \cdots v_j)$ due to the pumping Lemma.
    (We have $\delta_L(q_1,v_1 \cdots  v_i)=q_1$ for $i=0$.)
    Let $u_1 = v_1\cdots v_i, u_2 = v_{i+1}\cdots v_j$ and $u_3=v_{j+1}\cdots v_k$. Let $q_4 = \delta_L(q_1,u_1)$.

    We claim that $\cL_{q_2} \subseteq \cL_{q_4}$. The result then follows from $\cL_{q_2}=u_3^{-1}\cL_{q_2} \subseteq u_3^{-1}\cL_{q_4} = \cL_{q_3}$.
    To prove the claim, let $w=u_1u_2^N$ and $q_5=\delta_L(q_1,w^N)$. As $w \in \loopwords(q_2)$, we can use Corollary~\ref{cor:leftsync} to obtain $w^N \cL_{q_2} \subseteq \cL_{q_1}$.
     Together with $\cL_{q_5}=(w^N)^{-1}\cL_{q_1}$ this implies $\cL_{q_2} \subseteq \cL_{q_5}$.
    Furthermore, $u_2$ belongs to $\loopwords(q_5)$ because $L$ is aperiodic. To conclude the proof, we observe that $\cL_{q_5} \subseteq \cL_{q_4}$, by Corollary~\ref{cor:leftsync} with $q_5,q_4$ and $u_2$, and because $\delta_L(q_4,u_2^N)=q_4$ and $u_2 \in \loopwords(q_5)$.
\end{proof}

\begin{figure}[t] \centering
    \begin{tikzpicture}[->,>=latex,shorten >=.5pt,auto,node distance=1.3cm,
    inner sep = 0.7mm, initial text= $\cdots$]

    \node[state,initial,minimum size=17pt] (q1) at (-3,0) {$q_1$};
    \node[state,minimum size=17pt] (q2) at (-1,1) {$q_2$};
    \node[state,minimum size=17pt] (q3) at (.7,0.4) {$q_3$};
    \node[state,minimum size=17pt] (q4) at (-1.5,0) {$q_4$};
    \node[state,minimum size=17pt] (q5) at (.7,-.4) {$q_5$};

    \path (q1) edge  [->, loop above] node {$a..$} (q1);
    \path (q2) edge [->, loop above] node {$v_1,\ldots, v_N$} (q2);

    \path[draw,<-,decorate,
    decoration={snake,amplitude=.3mm,segment length=2.5mm,pre=lineto,pre
        length=5pt}] (q2) -- (q1);

    \path (q4) edge  node {$u_3$} (q3);

    \path (q4) edge [->, loop below] node {$u_2$} (q4);
    \path (q1) edge [->,swap] node {$u_1$} (q4);
    \path (q5) edge [->, loop right] node {$w, u_2$} (q5);
    \path (q4) edge [swap] node {$w^N$} (q5);

    \end{tikzpicture}
    \caption{Sketch of the proof of Lemma~\ref{lem:Bagan4}}%
    \label{fig:Bagan4}
\end{figure}

\begin{thm}
  Testing whether a regular language $L$ belongs to \ttract is
  \begin{enumerate}[(1)]
  \item NL-complete if $L$ is given by a DFA and
  \item PSPACE-complete if $L$ is given by an NFA or by a regular expression.
  \end{enumerate}
\end{thm}
\begin{proof}
	The proof is inspired by Bagan et al.~\cite{BaganBG-jcss20}. The
	upper bound for~(1) needs several adaptations, the lower bound for~(1)
	and the proof for (2) works exactly the same as
	in~\cite{BaganBG-arxiv12}, a preliminary version of~\cite{BaganBG-jcss20} (just replacing \sptract by \ttract).

  We first prove (1)\@. \mbox{W.l.o.g.}, we can assume that $L$ is given by the minimal
  DFA $A_L$, as testing Nerode-equivalence of two states is in NL\@.

By Lemma~\ref{lem:Bagan4}, we need to check for each
pair of states $q_1, q_2$ and symbol $a \in \Sigma$ whether
  \begin{enumerate}[(i)]
	\item $q_1 \leadsto q_2$;
	\item $\loopwords(q_1) \cap a \Sigma^* \neq \emptyset$; and
	\item $(\loopwords(q_2) \cap a \Sigma^*)^N\cL_{q_2} \setminus \cL_{q_1} = \emptyset$.
\end{enumerate}
Statements (i) and (ii) are easily verified using an NL algorithm
for transitive closure.  For (iii), we test emptiness of
$(\loopwords(q_2) \cap a \Sigma^*)^N\cL_{q_2} \setminus \cL_{q_1}$
using an NL algorithm for reachability in the product automaton of
$A_L$ with itself, starting in the state $(q_2,q_1)$. More
	precisely, the algorithm checks whether there does not exist a
	string that is in $\cL_{q_2}$, is not in $\cL_{q_1}$, starts with an
	$a$, and leaves the state $q_2$ (in the left copy of $A_L$) at least
	$N$ times with an $a$-transition.

 	 The remainder of the proof is from~\cite{BaganBG-arxiv12} and only
	included for self containedness.

	For the lower bound of (1), we give a reduction from the Emptiness problem.
	Let $L \subseteq \Sigma^*$ be
	an instance of Emptiness given by a DFA $A_L$. {W.l.o.g.} we assume that
	$\varepsilon \notin L$, since this can be checked in constant time.
	Furthermore, we assume that the symbol $1$ does not belong to $\Sigma$. Let
	$L' = 1^+L1^+$. A DFA $A_{L'}$ that recognizes $L'$ can be obtained from $A_L$
	as follows. We add a state $q_I$ that will be the initial state of $A_{L'}$.
	and a state $q_F$ that will be the unique final state of $A_{L'}$. The
	transition function $\delta_{L'}$ is the extension of $\delta_L$ defined as
	follows:
	\begin{itemize}
		\item $\delta_{L'}(q_I, 1) = q_I$ and $\delta_{L'}(q_I,a)=i_L$ for every
		symbol $a \in \Sigma$.
		\item For every final state $q \in F_L$, $\delta_{L'}(q,1) = q_F$.
		\item $\delta_{L'}(q_F, 1)=q_F$.
	\end{itemize}

    \noindent
	We will show that $L' \in \ttract$ if and only if $L$ is empty.
	If $L$ is empty, then $L' = \emptyset$ belongs to \ttract. For the other direction, assume that $L$ is
	not empty. Let $w \in L$. Then, for every $n \in \nat$, $1^n w 1^n \in L'$ and $1^n 1^n \notin L'$. Thus $L' \notin \ttract$.

	For the upper bound of (2), we first observe the following fact: Let $A,B$ be
	two problems such that $A \in$ NL and let $t$ be a reduction from $B$ to $A$
	that works in polynomial space and produces an exponential output. Then $B$
	belongs to PSPACE\@.
	Thus, we can apply the classical powerset
	construction for determinization on the NFA and use the upper bound from (1).

	For the lower bound of (2), we give a reduction from Universality. Let $L
	\subseteq \{0,1\}^*$ be an instance of Universality given by an NFA or a
	regular expression. Consider $L' = (0+1)^* a^* b a^* + La^*$ over the alphabet
	$\{0,1,a,b\}$.
	We show that $L = \{0,1\}^*$ if and only if $L' \in \ttract$.
	Our reduction associates $L'$ to $L$ and keeps the same
	representation (NFA or regular expression). If $L' = \{0,1\}^*$, then $L' = (0+1)^* a^* (b+\varepsilon)a^*$ and thus $L' \in \ttract$. Conversely, assume
	$L\neq \{0,1\}^*$. Let $w \in \{0,1\}^* \setminus L$. Then, for every $n \in \nat$,
	$wa^n b a^n \in L'$ and $wa^n a^n \notin L'$. Thus $L' \notin \ttract$.
\end{proof}

We wondered if, similarly to Theorem~\ref{theo:downwardclosed}, it could be the
case that languages closed under \lsla are always regular, but this is not the
case. %
For example, the (infinite) Thue-Morse
word~\cite{Thue1906,Morse1921} has no subword that is a cube (i.e., no subword
of the form $w^3$)~\cite[Satz~6]{Thue1906}. The language containing all
prefixes of the Thue-Morse word thus trivially is closed under \lsla (with
$i=3$), yet it is not regular.

We now give some closure properties of \ctract and \ttract. We note that Bagan et al.~\cite{BaganBG-jcss20} already observed that \sptract is closed under finitie unions, intersections, and reversal.

\begin{lem}\label{lem:closure}
  Both classes \ctract and \etract are closed under
  \emph{(i)} finite unions,
  \emph{(ii)} finite intersections,
  \emph{(iii)} reversal,
  \emph{(iv)} left and right quotients,
  \emph{(v)} inverses of non-erasing morphisms,
  \emph{(vi)} removal and addition of individual strings.
\end{lem}
\begin{proof}
 The closure properties (i) to (vi) follow immediately from Observation~\ref{obs:ne-varieties}, i.e., that \sptract and \ttract are ne-varieties, see~\cite{Pin-handbook97,PinS-ita05}.
\end{proof}
\begin{lem}
    The classes \ctract and \etract are not closed under complement.
\end{lem}
\begin{proof}
    Let $\Sigma = \{a,b\}$. The language of the expression $b^*$ clearly is in \ctract and \ttract.
    Its complement is the language $L$ containing all words with at least one $a$. It can be described by the regular expression $\Sigma^* a \Sigma^*$.
    Since $b^i a b^i \in L$ for all $i$, but $b^i b^i \notin L$ for any $i$, the language $L$ is neither in \ctract nor in \ttract.
\end{proof}

It is an easy consequence of Lemma~\ref{lem:closure}~(vi)
that for regular
languages outside of $\ctract$ or $\ttract$ there do not exist best lower or upper approximations.
\begin{cor}
  Let $\cC \in \{\ctract,\ttract\}$. For every regular language $L$ such
  that $L\notin \cC$ and
  \begin{itemize}
    \item for every upper approximation $L''$ of $L$ (i.e.,
  $L \subsetneq L''$) with $L'' \in \cC$ it holds that there exists a
  language $L' \in \cC$ with $L \subsetneq L' \subsetneq L''$;
    \item for every lower approximation $L''$ of $L$ (i.e., $L'' \subsetneq L$) it holds that there exists a
      language $L' \in \cC$ with $L'' \subsetneq L' \subsetneq L$.
    \end{itemize}
\end{cor}

\noindent
The corollary implies that Angluin-style learning of languages in \ctract or
\ttract is not possible. However, learning algorithms for single-occurrence
regular expressions (SOREs) exist~\cite{BexNSV-tods10} and can therefore be useful for
an important subclass of \ttract.

\section{Enumeration}\label{sec:enum}
In this section we state that---using the algorithm from
Theorem~\ref{theo:trichotomy}---the enumeration result
from~\cite{Yen-ms71} transfers to the setting of enumerating trails
matching $L$.
\begin{thm}
  Let $L$ be a regular language, $G$ be a multigraph and $(s,t)$ a pair of nodes in
  $G$. If NL $\neq$ NP, then one can enumerate trails from $s$ to $t$ that match
  $L$ in polynomial delay in data complexity if and only if $L \in \etract$.
\end{thm}
\begin{proof}[Proof sketch]
  The algorithm is an adaptation of Yen's algorithm~\cite{Yen-ms71} that
  enumerates the $k$ shortest simple paths for some given number $k$, similar to
  what was done by Martens and Trautner~\cite{MartensT-tods19}.
  It uses the algorithm %
  from Corollary~\ref{cor:output-shortest-paths}
  as a subprocedure.
\end{proof}
Notice that we cannot simply use the line graph construction and
solve this problem for simple paths since the class of regular languages that is
tractable for simple paths is a strict subset of \etract. So this method would
not, for example, solve the problem for $L = (ab)^*$.

Instead, we can change Yen's algorithm~\cite{Yen-ms71} to work with trails
instead of simple paths. The changes are straightforward: instead of deleting
nodes, we only delete edges. In Algorithm~\ref{alg:yen} we changed Yen's algorithm to enumerate
$L$-labeled trails. Note that we only need $L \in \etract$ to ensure that
the subroutines in lines~\ref{alg:yen:3} and~\ref{alg:yen:12} are in polynomial time.

\begin{algorithm}[t] %
  \DontPrintSemicolon%
  \KwIn{Multigraph $G=(V,E, \edge)$, nodes $s,t \in V$, a language $L \in \etract$}
  \KwOut{All trails from $s$ to $t$ in $G$ that match $L$ under bag semantics}
  $A \gets \emptyset$ \Comment*[r]{\textrm{$A$ is the set of trails already written to output}}
  $B \gets \emptyset$ \Comment*[r]{\textrm{$B$ is a set of trails from $s$ to $t$ matching  $L$}}
  $p \gets$ a shortest trail from $s$ to $t$ matching $L$\label{alg:yen:3} \Comment*[r]{\textrm{$p \gets \bot$ if no such trail exists}}
  \While{$p \neq \bot$}{
    \textbf{output} $p$\label{alg:yen:output1}\; Add $p$ to $A$ \;
    \For{$i = 0$ to $|p|$\label{alg:yen:for1} }{
      $G' \gets (V, E', \edge|_{E'})$ with $E'=E\setminus E(p[1,i])$
      \Comment*[r]{\textrm{Remove edges used in $p[1,i]$}}
      $S \gets \{e \in E \mid p[1,i] \cdot e \text{ is a prefix of a trail in } A\}$\label{alg:yen:S} \;
      $p_1 \gets$ a shortest trail from $\dest(p[1,i])$ to $t$ in $G'$ that matches $((\lab(p[1,i]))^{-1} L) \setminus\{\varepsilon\}$ and does not start with an edge from $S$\label{alg:yen:12}\;
      Add $p[1,i]\cdot p_1$ to $B$
    }
    $p \gets$ a shortest trail in $B$ \Comment*[r]{\textrm{$p \gets \bot \text{ if } B = \emptyset$}} %
    Remove $p$ from $B$
  }
  \caption{Yen's algorithm changed to work with trails on multigraphs}%
  \label{alg:yen}
\end{algorithm}

\paragraph*{Explanation of Algorithm~\ref{alg:yen}}
In the algorithm, $p[1,i]$ denotes the prefix of $p$ containing exactly $i$ edges and $\dest(p)$ denotes the last node of $p$.
In line~\ref{alg:yen:3}, we can use the algorithm explained in the
proof of Theorem~\ref{theo:trichotomy} (more concretely, Corollary~\ref{cor:output-shortest-paths})
to find a $L$-labeled trail
from $s$ to $t$ in $G$ in NL if one exists. In the for-loop in
line~\ref{alg:yen:for1} we use quotients of the last trail written
to the output to find new candidates.
Intuitively, for all
$i\in \nat$, we regard all paths that share the prefix of length
exactly $i$ with the last path and do not share a prefix of length
$i+1$ with any path outputted so far. In line~\ref{alg:yen:12}, we
search for a suffix to the prefix $p[1,i]$ by again using the algorithm
explained in the proof of Theorem~\ref{theo:trichotomy}. We recall
that $\ttract$ is closed under left derivatives and removal of individual strings, see Lemma~\ref{lem:closure}, i.e.,
$(\lab(p[1,i]))^{-1}L \setminus \{\varepsilon\}$ is in $\ttract$.  However, to prevent finding a
trail that was already in the output, we do not allow the suffix to start
with some edge from $S$. We note that the algorithm from Corollary~\ref{cor:output-shortest-paths} can be easily modified to check for this
additional condition. We repeat this procedure with all trails in $B$,
until we do not find any new trails.

\paragraph*{Set vs.\ Bag Semantics in Multigraphs}
Let us consider a small multigraph $G=(V,E,\edge)$ with two nodes $\{v_1, v_2\}$
and two edges $e_1,e_2$ with $\edge(e_1)=\edge(e_2)=(v_1,a,v_2)$. If we want to
enumerate all paths from $v_1$ to $v_2$ that match $a$, how many paths should we
get? Under set semantics, we will only obtain a single answer, whereas, under
bag semantics, we will consider $e_1$ and $e_2$ as different edges and therefore
return two answers. Algorithm~\ref{alg:yen} enumerates trails according to bag
semantics.
This is because all edges are considered to be different.
Algorithm~\ref{alg:set-semantics} enumerates trails according to set semantics.

\begin{algorithm}[t] %
	\DontPrintSemicolon%
	\KwIn{Multigraph $G=(V,E, \edge)$, nodes $s,t \in V$, a language $L \in \etract$}
	\KwOut{All trails from $s$ to $t$ in $G$ that match $L$ under bag semantics}
	$A \gets \emptyset$ \Comment*[r]{\textrm{$A$ is the set of trails already written to output}}
	$B \gets \emptyset$ \Comment*[r]{\textrm{$B$ is a set of trails from $s$ to $t$ matching  $L$}}
	$p \gets$ a shortest trail from $s$ to $t$ matching $L$  \Comment*[r]{\textrm{$p \gets \bot$ if no such trail exists}}
	$j \gets 0$ \Comment*[r]{\textrm{tells where the deriviation should start}}
	\While{$p \neq \bot$}{
		\textbf{output} $p$\; Add $p$ to $A$ \;
		\For{$i = j$ to $|p|$\label{alg:set-for-loop} }{
			$G' \gets (V, E', \edge|_{E'})$ with $E'=E\setminus E(p[1,i])$
			\Comment*[r]{\textrm{Remove edges used in $p[1,i]$}}
			$S \gets \{e' \in E \mid \exists e \text{ with } \edge(e) = \edge(e') \text{
				and } p[1,i] \cdot e \text{ is a prefix of a trail in } A\}$\;
			$p_1 \gets$ a shortest trail from $\dest(p[1,i])$ to $t$ in $G'$ that matches $((\lab(p[1,i]))^{-1} L) \setminus\{\varepsilon\}$ and does not start with an edge from $S$ \;
			Add $(p[1,i]\cdot p_1,i)$ to $B$
		}
		$(p,j) \gets$ a pair from $B$ with $p$ being a shortest trail \Comment*[r]{\textrm{$p \gets \bot \text{ if } B = \emptyset$}} %
		Remove $(p,j)$ from $B$
	}
	\caption{Yen's algorithm for trails with set semantics}%
	\label{alg:set-semantics}
\end{algorithm}

The changes between Algorithm~\ref{alg:yen} and Algorithm~\ref{alg:set-semantics} are the following:
\begin{enumerate}[(1)]
	\item The set $S$ is computed differently. In Algorithm~\ref{alg:yen} the set $S$ contains exactly the edges that were already used to continue the path $p[1,i]$, whereas in Algorithm~\ref{alg:set-semantics} the set $S$ contains  all edges whose origin, destination, and label are identical to some edge already used to continue the path $p[1,i]$.
	\item In Algorithm~\ref{alg:yen} the set $B$ can contain paths that are identical under set semantics, while this is forbidden Algorithm~\ref{alg:set-semantics}. One possibility to realize this is in Algorithm~\ref{alg:set-semantics} is to use
    Lawler's~\cite{Lawler-MS72} extension of Yen's algorithm. Lawler observed
    that if a path $p' = p[1,i] \cdot p_1$ was added to $A$, in the next
    iteration it is sufficent to start the derivation from $i$, as the
    derivatives of $p[1,i]$ have already been added to $B$. This change impacts the for-loop in line~\ref{alg:set-for-loop} of Algorithm~\ref{alg:set-semantics}.
\end{enumerate}
We note that Lawler's extension of Yen's algorithm can also be used under bag
semantics to improve the running time, as observed by Lawler~\cite{Lawler-MS72}.
The following example shows that (1) alone is not
sufficient to enumerate multigraphs under set semantics because it would be
possible that different paths are added to $B$ which are identical under set
semantics, but different under bag semantics.
 \begin{exa}\label{example:set-semantics}
   Consider the graph illustrated in Figure~\ref{fig:enum-set-semantics} with
   nodes $\{s,v,t\}$ and edges $\{e_1, \ldots, e_5\}$.
   Assume $S$ in line~\ref{alg:yen:S} in Algorithm~\ref{alg:yen} is calculated
   as in (1). If the algorithm starts with the $ab$-path $e_1 e_2$, it can add
   one $db$-path $e_4 e_2$ and the $ab$-path $e_1 e_3$ to $B$. If it picks the
   $ab$-path $e_1 e_3$ from $B$ in the next step, the algorithm might add
   the other $db$-path $e_5 e_2$ to $B$. Thus $B = \{ e_4e_2, e_5e_2\}$ contains
   two paths that are identical under set semantics.

   \begin{figure}[t] %
   	\centering {
   		\begin{tikzpicture}[auto,>=latex,->]%
   			\node (s) at (0,0) {};
   			\node (v1) at (3,0) {};
   			\node (t) at (6,0) {};

   			\foreach \i in {s,v1,t}{
   				\fill (\i) circle (2pt);
   			}

   			\draw (s) coordinate[label={left:$s$}];
   			\draw (t) coordinate[label={right:$t$}];

   			\path (s) edge [bend left = 50] node {$a/e_1$} (v1);
   			\path (s) edge node {$d/e_4$} (v1);
   			\path (s) edge [bend right =50] node[below] {$d/e_5$} (v1);

   			\path (v1) edge [bend left = 50] node {$b/e_2$} (t);
   			\path (v1) edge [bend right = 50] node[below] {$c/e_3$} (t);
   	\end{tikzpicture}
    }
   	\caption{Example graph for enumeration under set semantics. Edges are
      annotated with label and edge identifier.
   	}%
   	\label{fig:enum-set-semantics}
   \end{figure}
 \end{exa}

\section{Conclusions and Lessons Learned}\label{sec:conclusion}
We have defined the class \ttract of regular languages $L$ for which finding trails
in directed graphs that are labeled with $L$ is tractable iff \nlogspace $\neq$
NP\@. We have investigated \ttract in depth in terms of closure properties,
characterizations, and the recognition problem, also touching upon the closely
related class \sptract (for which finding \emph{simple paths} is tractable) when relevant.

In our view, graph database manufacturers can have the following trade-offs in
mind concerning simple path (\sptract) and trail semantics (\ttract) in database
systems:
\begin{itemize}
\item $\sptract \subsetneq \ttract$, that is, there are strictly more languages
  for which finding regular paths under trail semantics is tractable than under
  simple path semantics. Some of the languages in \ttract but outside \sptract
  are of the form $(ab)^*$ or $a^*bc^*$, which were found to be relevant in
  several application scenarios involving network problems, genomic datasets,
  and tracking provenance information of food products~\cite{PetraSelmerPersonal} and appear in query
  logs~\cite{BonifatiMT-vldb17,BonifatiMT-www19}.
\item Both \sptract and \ttract can be syntactically characterized but,
  currently, the characterization for \sptract (Section 7 in~\cite{BaganBG-jcss20}) is simpler than the one for
  \ttract. This is due to the fact that connected components for
  automata for languages in \ttract can be much more complex than for automata
  for languages in \ctract.
\item On the other hand, the \emph{single-occurrence} condition, i.e., each
  alphabet symbol occurs at most once, is a sufficient condition for regular
  expressions to be in \ttract. This condition is trivial to check and also captures
  languages outside $\sptract$ such as $(ab)^*$ and $a^*bc^*$. Moreover, the
  condition seems to be useful:  we analyzed the 50 million
  RPQs found in the logs of~\cite{BonifatiMT-www18} and discovered that over
  99.8\% of the RPQs are single-occurrence.

\item In terms of closure properties, learnability, or complexity of testing if a given regular
  language belongs to \ctract or \ttract, the classes seem to behave the same.

\item The tractability for the decision version of RPQ evaluation can be lifted
  to the enumeration problem, in which case the task is to output matching paths
  with only a polynomial delay between answers.
\end{itemize}

\noindent
As an open question remains the trichotomy for 2RPQs, that is, when we allow
RPQs to follow a directed edge also in its reverse direction. We briefly discuss
why this is challenging. Let us denote by $\hat{\ } a$ the backward navigation
of an edge labeled $a$. Then, the case of ordinary RPQs can be seen as a special
case of 2RPQs on undirected graphs: it only has bidirectional navigation of the form $(a +
\hat{\ } a)$. It has been open since 1991 whether evaluating $(aaa)^*$
on undirected graphs is in P or NP-complete~\cite{ArkinPY-jacm91}.

\subparagraph{Acknowledgments} This work was supported by DFG grant MA 4938/4-1. 
 We thank the participants of Shonan meeting
No.~138 (and Hassan Chafi in particular), who provided significant inspiration
for the first paragraph in the Introduction, Jean-\'Eric Pin for pointing us
to positive $C_\text{ne}$-varieties of languages, and Charles Paperman for useful comments and pointing out that $\ttract \subsetneq \Pi[<,+1]$ (see Theorem~\ref{theo:between-pi1s}). We also thank Jean-\'Eric Pin and Luc
Segoufin for their help with the proof with a weaker statement.
Furthermore, we want to thank the anonymous reviewers for valuable comments.

\bibliographystyle{alphaurl}
 \bibliography{references}

\end{document}